\documentclass[aps,prd,showpacs,showkeys,amsmath,amssymb]{revtex4}
\usepackage{graphicx}
\usepackage{bm}
\begin{document}

\title{Heavy and Light Pentaquark Chiral Lagrangian}
\author{Y.-R. Liu, A. Zhang, P.-Z. Huang, W.-Z. Deng,
X.-L. Chen}
\affiliation{%
Department of Physics, Peking University, BEIJING 100871, CHINA}
\author{Shi-Lin Zhu}
\email{zhusl@th.phy.pku.edu.cn}
\affiliation{%
Department of Physics, Peking University, BEIJING 100871, CHINA}
\affiliation{%
COSPA, Department of Physics, National Taiwan University, Taipei
106, Taiwan, R.O.C.}

\date{\today}

\begin{abstract}

Using the $SU(3)$ flavor symmetry, we construct the chiral
Lagrangians for the light and heavy pentaquarks. The correction
from the nonzero quark is taken into account perturbatively. We
derive the Gell-Mann$-$Okubo type relations for various pentaquark
multiplet masses and Coleman-Glashow relations for anti-sextet
heavy pentaquark magnetic moments. We study possible decays of
pentaquarks into conventional hadrons. We also study the
interactions between and within various pentaquark multiplets and
derive their coupling constants in the symmetry limit. Possible
kinematically allowed pionic decay modes are pointed out.

\end{abstract}


\maketitle

\pagenumbering{arabic}

\section{Introduction}\label{sec1}

Since LEPS Collaboration announced the discovery of the exotic
baryon with very narrow width $\Theta^+(1540)$ \cite{leps}, many
other groups have claimed the observation of this state
\cite{diana,clas,saphir,itep,clasnew,hermes,svd,cosy,
Yerevan,zeus,forzeus}. NA49 observed a new pentaquark
$\Xi^{--}(1862)$ \cite{na49}, which needs confirmation from other
groups \cite{doubt}. Recently, H1 Collaboration claimed the
discovery of a heavy pentaquark around 3099 MeV with the quark
content $udud\bar{c}$ \cite{H1}. It is interesting to note that
several groups reported negative results \cite{bes,hera-b,rhic}.

There is preliminary evidence that the $\Theta^+$ is an iso-scalar
because no enhancement was observed in the $pK^+$ invariant mass
distribution \cite{saphir,clasnew,hermes,forzeus}. Most of the
theoretical models assume that $\Theta^+$ is in $SU(3)_f$
${\bf\bar{10}}$ representation.

The parity of $\Theta$ pentaquark remains unknown. Theoretical
approaches advocating positive parity include the chiral soliton
model (CSM) \cite{diak}, the diquark model \cite{jaffe}, some
quark models \cite{lipkin,positive,carl,cohen,liu}, a lattice
calculation \cite{chiu}. On the other hand, some other theoretical
approaches tend to favor negative parity such as two lattice QCD
simulations \cite{lattice,sasaki}, QCD sum rule approaches
\cite{zhu,qsr}, several quark model study \cite{zhang,carlson,wu},
and proposing stable diamond structure for $\Theta^+$ \cite{song}.

The narrow width of $\Theta$ pentaquark is another puzzle. All the
experiments can only determine the upper bound of the pentaquark
widths up to the detector resolution. The reanalysis of previous
pion kaon scattering data indicates the decay width of $\Theta^+$
should be one or two MeV or less \cite{width}, which makes the
theoretical interpretation very difficult.

There have appeared several attempts to explain the narrow width.
One possibility is the mismatch between the spin-flavor wave
functions of the initial and final state when $\Theta$ pentaquark
decays through the fall-apart mechanism
\cite{maltman,carl,close,buccella}.

Another possible interpretation of the narrow width puzzle is the
possible mismatch between the spatial wave functions of final and
initial states \cite{song}. The reason is simple. The $\Theta^+$
pentaquark with the stable diamond structure and bound by
non-planar flux tubes is hard to decay into hadrons bound by
planar flux tubes \cite{song}. But this scheme has not been
studied quantitatively.

In the chiral soliton model, the narrowness of $\Theta^+$ results
from the cancellation of the coupling constants at different $N_c$
orders \cite{ellis}. It is suggested that one of two nearly
degenerate pentaquarks sharing the same dominant decay mode can be
arranged to decouple from the decay channel after diagonalizing
the mixing mass matrix via kaon nucleon loop \cite{width2}.

Recently heavy pentaquarks have received much attention
\cite{jaffe,lipkin,huntheavy,dudek,bicudo,sasaki,leandri,stancu,kingman,
huang2,wise,he,chiu2,cheng}. In the heavy quark limit, the heavy
anti-quark decouples and acts as a spectator. The pentaquark
system simplifies significantly. In fact, the heavy pentaquark
system can be used as a test-ground of the various models
developed for the light pentaquarks.

Model calculation has shown that the anti-decuplet and the
even-parity pentaquark octet lie close to each other and ideal
mixing occurs if quantum number allows \cite{jaffe}. The
odd-parity pentaquark nonet is several hundred MeV lower than the
anti-decuplet and even-parity octet. Strong transitions between
different pentaquark multiplets may occur \cite{zhanga}.

At present the underlying dynamics which binds four quarks and one
anti-quark into a narrow resonance above threshold is still a
mystery. We will explore the strong interactions between
pentaquark multiplets using the $SU(3)$ flavor symmetry as the
guide. Chiral Lagrangians have been used to study the strong decay
modes of pentaquarks \cite{zhanga,lee,he,ko,mehen}.

In Section \ref{sec2}, we will construct the chiral Lagrangian
involving light and heavy pentaquark multiplets. Then we discuss
the mass splitting from the current quark mass correction within
the same multiplet.  In Section \ref{sec4}, we derive the coupling
constants of the pentaquark interactions and discuss possible
strong decay modes. The final section is a short discussion.

\section{Chiral Lagrangian}\label{sec2}

\subsection{Notation}

The approximate chiral symmetry and its spontaneous breaking have
played an important role in hadron physics. Through the nonlinear
realization of spontaneous chiral symmetry breaking, we may study
the interaction between the chiral field and hadrons, which always
involves the derivative of the chiral field. The nonzero current
quark mass breaks the chiral symmetry explicitly. These
corrections are taken into account perturbatively together with
the chiral loop correction. Generally speaking, chiral symmetry
provides a natural framework to organize the hadronic strong
interaction associated with the light quarks.

In writing down the pentaquark chiral Lagrangians, we follow the
standard notation in the chiral perturbation theory. First the
eight Goldstone bosons are introduced exponentially. We use the
short-hand notation $\pi$ to denote them.
\begin{equation}
\Sigma\equiv\xi^2\equiv\exp({2i\pi\over F_\pi})\; ,
\end{equation}
\begin{eqnarray}
\pi &=&\left(\begin{array}{ccc}
\frac{\pi^0}{\sqrt{2}}+\frac{\eta_0}{\sqrt{6}}&\pi^+&K^+\\
 \pi^-&-\frac{\pi^0}{\sqrt{2}}+\frac{\eta_0}{\sqrt{6}}&K^0\\
K^-&\bar{K}^0&-\frac{2\eta_0}{\sqrt{6}} \end{array}\right)\; ,
\end{eqnarray}
where $F_\pi = 92.4$ MeV is the pion decay constant.

Under the $SU(3)_L\times SU(3)_R$ chiral transformation,
$\Sigma(x)$ and $\xi(x)$ transform as
\begin{eqnarray}
\Sigma(x)&\rightarrow& L\Sigma(x) R^\dagger, \nonumber\\
\xi(x)&\rightarrow& L\xi(x) U^\dagger(x)=U(x)\xi(x)R^\dagger
\end{eqnarray}
where $L\in SU(3)_L$, $R\in SU(3)_R$, $U(x)$ is a non-linear
function of $\pi(x)$ and $L, R$.

The chiral connection $V_\mu$ and the axial vector field $A_\mu$
are defined as
\begin{eqnarray}
V_{\mu} = \frac{1}{2}(\xi^\dagger\partial_\mu \xi + \xi
\partial_\mu  \xi^\dagger ),
\nonumber\\
A_{\mu} = \frac{i}{2}(\xi^\dagger\partial_\mu \xi - \xi
\partial_\mu \xi^\dagger).
\end{eqnarray}
The vector $V_\mu$ and axial vector $A_\mu$ transform under chiral
$SU(3)$ as
\begin{eqnarray}
 V_\mu &\to&  U V_\mu U^{\dagger} + U
\partial_\mu U^{\dagger},\nonumber\\
A_\mu &\to&  U A_\mu U^{\dagger}.
\end{eqnarray}

With the chiral connection, we can construct the chirally
covariant derivative ${\cal D}_\mu$. For the matter field $\phi$
which is in the fundamental representation, we have
\begin{eqnarray}
{\cal D}_{\mu} \phi = \left( \partial_\mu +V_\mu\right) \phi ,
\nonumber\\
{\cal D}_\mu \to  U {\cal D}_\mu U^{\dagger}.
\end{eqnarray}

For the matter field in the adjoint representation like the
nucleon octet $B$, we have
\begin{equation}
{\cal D}_{\mu} B = \partial_\mu B +\left[ V_\mu, B \right]
\end{equation}
where the octet baryon field reads
\begin{eqnarray}
(B_j^i)&=&\left(\begin{array}{ccc}
\frac{\Sigma^0}{\sqrt{2}}+\frac{\Lambda}{\sqrt{6}}&\Sigma^+&p\\
 \Sigma^-&-\frac{\Sigma^0}{\sqrt{2}}+\frac{\Lambda}{\sqrt{6}}&n\\
\Xi^-&\Xi^0&-\frac{2\Lambda}{\sqrt{6}} \end{array}\right)\; .
\end{eqnarray}

For the $\Delta^{++}$ decuplet, the chirally covariant derivative
reads
\begin{eqnarray}
{\cal D}_\mu D^{ijk}=\partial_\mu D^{ijk} + V^i_{\mu,\,a} D^{ajk}
+ V^j_{\mu,\,a} D^{iak} +V^k_{\mu,\,a} D^{ija}.
\end{eqnarray}

\subsection{Matter fields}

In Jaffe and Wilczek's diquark model \cite{jaffe}, the color wave
function of the two diquarks within the pentaquark must be
antisymmetric $\bf{3}_C$. In order to get an exotic anti-decuplet,
the two scalar diquarks combine into the symmetric SU(3)
$\bf{\bar{6}_F}$ : $[ud]^2$, $[ud][ds]_+$, $[su]^2$, $[su][ds]_+$,
$[ds]^2$, and $[ds][ud]_+$. Bose statistics demands symmetric
total wave function of the diquark-diquark system, which leads to
the antisymmetric spatial wave function with one orbital
excitation. The resulting anti-decuplet $P_{ijk}$ and octet
pentaquarks $O_{1j}^i$ have $J^P={1\over 2}^+, {3\over 2}^+$.

Our discussion makes use of the flavor symmetry only. So the
results are valid for both $J^P={1\over 2}^+$ and $ {3\over 2}^+$.
We use $J^P={1\over 2}^+$ case to illustrate the formalism.

The members of pentaquark anti-decuplet are $P_{333}=\Theta^+$, $
P_{133}=\frac{1}{\sqrt{3}}N^0_{10}$,
$P_{233}=-\frac{1}{\sqrt{3}}N^+_{10}$, $P_{113}
=\frac{1}{\sqrt{3}}\Sigma^-_{10}$,
$P_{123}=-\frac{1}{\sqrt{6}}\Sigma^0_{10}$,
$P_{223}=\frac{1}{\sqrt{3}}\Sigma^+_{10}$,
$P_{111}=\Xi^{--}_{10}$, $P_{112} =-\frac{1}{\sqrt{3}}\Xi^-_{10}$,
$P_{122}=\frac{1}{\sqrt{3}}\Xi^0_{10}$ and $P_{222}=-\Xi^+_{10}$.

Later, we pointed out \cite{zhanga} that lighter pentaquarks can
be formed if the two scalar diquarks are in the antisymmetric
$SU(3)_F$ $\bf 3$ representation: $[ud][su]_-$, $[ud][ds]_-$, and
$[su][ds]_-$, where
$[q_1q_2][q_3q_4]_-=\sqrt{\frac{1}{2}}([q_1q_2][q_3q_4]-[q_3q_4][q_1q_2])$.
No orbital excitation is needed to ensure the symmetric total wave
function of two diquarks since the spin-flavor-color part is
symmetric. The total angular momentum of these pentaquarks is
$\frac{1}{2}$ and the parity is negative. There is no accompanying
$J={3\over 2}$ multiplet. The two diquarks combine with the
antiquark to form a $SU(3)_F$ pentaquark octet $O_{1j}^i$ and
singlet $\Lambda_1$.

Replacing the light anti-quark by one anti-charm or anti-bottom
quark in Jaffe and Wilczek's model leads to one even parity
anti-sextet $S_{ij}$ \cite{jaffe,huang2} and one odd parity
triplet $T^i$ \cite{kingman,wise}. In Karliner and Lipkin's
diquark tri-quark model, there is an additional even parity heavy
pentaquark triplet \cite{huang2}. In the following discussion, we
only make use of the flavor symmetry to write down the chiral
Lagrangian. Hence the results are not limited to Jaffe and
Wilczek's model only. The heavy pentaquark multiplets are
\begin{eqnarray}
(S_{ij}^c)&=&\left(\begin{array}{ccc}
\Xi^{--}_{5c}&-\frac{1}{\sqrt{2}}{\Xi^-_{5c}}&\frac{1}{\sqrt{2}}{\Sigma^-_{5c}}\\
-\frac{1}{\sqrt{2}}{\Xi^-_{5c}}&\Xi^0_{5c}&-\frac{1}{\sqrt{2}}{\Sigma^0_{5c}}\\
\frac{1}{\sqrt{2}}{\Sigma^-_{5c}}&-\frac{1}{\sqrt{2}}{\Sigma^0_{5c}}&\Theta^0_{5c}\end{array}\right),\nonumber
\\
(S_{ij}^b)&=&\left(\begin{array}{ccc}
\Xi^{-}_{5b}&-\frac{1}{\sqrt{2}}{\Xi^0_{5b}}&\frac{1}{\sqrt{2}}{\Sigma^0_{5b}}\\
-\frac{1}{\sqrt{2}}{\Xi^0_{5b}}&\Xi^+_{5b}&-\frac{1}{\sqrt{2}}{\Sigma^+_{5b}}\\
\frac{1}{\sqrt{2}}{\Sigma^0_{5b}}&-\frac{1}{\sqrt{2}}{\Sigma^+_{5b}}&\Theta^+_{5b}\end{array}\right),\nonumber
\\
(T^i_c)&=&\left(\begin{array}{c} \Sigma^{\prime0}_{5c} \\ \Sigma^{\prime-}_{5c}\\
\Xi^{\prime-}_{5c}\end{array}\right),\\
(T^i_b)&=&\left(\begin{array}{c} \Sigma^{\prime+}_{5b} \\ \Sigma^{\prime0}_{5b}\\
\Xi^{\prime0}_{5b}\end{array}\right).
\end{eqnarray}

In writing down the chiral Lagrangians, we need the pseudoscalar
heavy meson triplet $\bar{Q}^i$ in the fundamental representation:
\begin{eqnarray}
({Q}_i)&=&\left(\begin{array}{ccc}Q\bar{u}, &Q\bar{d}, &
Q\bar{s}\end{array}\right) \; .
\end{eqnarray}

Under chiral transformation, the matter fields transform as
\begin{eqnarray}
B^i_j&\rightarrow&U^i_a B^a_b U^{\dagger b}_j,\nonumber\\
D^{ijk}&\rightarrow& U^i_a U^j_b U^k_c D^{abc},\nonumber\\
O^i_{1j}&\rightarrow&U^i_a O^a_{1b} U^{\dagger b}_j,\nonumber\\
O^i_{2j}&\rightarrow&U^i_a O^a_{2b} U^{\dagger b}_j,\nonumber\\
\Lambda_1 &\rightarrow& \Lambda_1 ,\nonumber\\
P_{ijk}&\rightarrow&P_{abc}U^{\dagger a}_i U^{\dagger b}_j U^{\dagger c}_k,\nonumber\\
\bar{Q}^i&\rightarrow& U^i_a \bar{Q}^a, \nonumber\\
S_{ij}&\rightarrow&S_{ab}U^{\dagger a}_i U^{\dagger b}_j,\nonumber\\
T^i&\rightarrow& U^i_a T^a .
\end{eqnarray}

The chirally covariant derivatives of these matter fields have the
same transformation as the matter fields. They are
\begin{eqnarray}
{\cal D}_\mu B^i_j&=&\partial_\mu B^i_j+V^i_{\mu,\,a}B^a_j-B^i_a V^a_{\mu,\,j}\;,\nonumber\\
{\cal D}_\mu O^i_{1j}&=&\partial_\mu O^i_{1j}+V^i_{\mu,\,a}O^a_{1j}-O^i_{1a} V^a_{\mu,\,j}\;,\nonumber\\
{\cal D}_\mu O^i_{2j}&=&\partial_\mu O^i_{2j}+V^i_{\mu,\,a}O^a_{2j}-O^i_{2a} V^a_{\mu,\,j}\;,\nonumber\\
{\cal D}_\mu P_{ijk}&=&\partial_\mu P_{ijk} + P_{ija}V^{\dagger
a}_{\mu,\,k} + P_{iak}V^{\dagger a}_{\mu,\,j} + P_{ajk}V^{\dagger
a}_{\mu,\,i}\;,\nonumber\\
{\cal D}_\mu\bar{Q}^i &=& \partial_\mu\bar{Q}^i +
V^i_{\mu,\,j}\bar{Q}^j \;,\nonumber\\
{\cal D}_\mu S_{ij}&=&\partial_\mu S_{ij} + S_{ia}V^{\dagger
a}_{\mu,\,j} + S_{aj}V^{\dagger a}_{\mu,\,i} \;,\nonumber\\
{\cal D}_\mu T^i &=&\partial_\mu T^i+V^i_{\mu,\,j} T^j .
\end{eqnarray}

The current quark mass matrix $m=\textrm{diag}(\hat{m}, \hat{m},
m_s)$ transforms as $m \rightarrow L m R^\dagger=R m L^\dagger$
under $SU(3)_L\times SU(3)_R$ chiral transformation, where we have
ignored the isospin breaking effect and adopt $m_u=m_d=\hat{m}$.
Hence, the following combination of $m$ and $\xi$ transforms as
the matter field:
\begin{eqnarray}
(\xi m\xi+\xi^\dagger m\xi^\dagger)&\rightarrow&U(\xi
m\xi+\xi^\dagger m\xi^\dagger)U^\dagger.
\end{eqnarray}

\subsection{Mass, kinetic term and interaction with the chiral field}

With these matter fields and their corresponding transformations
under chiral transformation, we first write down the chiral
Lagrangian involving mass term, kinematic terms, and interaction
terms between the matter field and the chiral field.
\begin{equation}
\mathcal{L} = \mathcal{L}_{\Sigma} + \mathcal{L}_{B}
+\mathcal{L}_P +\mathcal{L}_{O_1}+\mathcal{L}_{O_2}+ {\mathcal
L}_{\Lambda_1} +{\mathcal L}_Q+ \mathcal{L}_S
+\mathcal{L}_T+\mathcal{L}_{int},
\end{equation}
where
\begin{subequations}
\begin{eqnarray}
{\cal L}_{\Sigma} & = & {F_\pi^2 \over 4}\,\textrm{Tr} \big[
{\partial}_\mu \Sigma^{\dagger} {\partial}^{\mu} \Sigma -2\mu m (
\Sigma + \Sigma^{\dagger} ) \big],
\\
{\cal L}_{B} & = & \textrm{Tr}\,\overline{B} (i
\mathcal{D}\!\!\!\!/-m_B) B \nonumber\\
&&- D_B\,\textrm{Tr}\,\overline{B} \gamma^\mu\gamma_5 \{A_\mu,B\}
 - F_B\,\textrm{Tr}\,\overline{B} \gamma^\mu\gamma_5 [A_\mu,B],
\\
{\cal L}_D&=&\bar{D}(i {\cal D}\!\!\!\!/-m_D)D + {\cal G}_D
\overline{D}
\gamma_5 A \!\!\!/ D,\\
{\cal L}_P & = & \overline{P}(i
\mathcal{D}\!\!\!\!/-m_P)P+\mathcal{G}_{P} \overline{P}\gamma_5
A\!\!\!/ P ,
\\
{\cal L}_{O_1} & = & \textrm{Tr}\,\overline{O_1} (i
\mathcal{D}\!\!\!\!/-m_{O_1}) O_1 \nonumber\\
&&- D_{O_1}\,\textrm{Tr}\,\overline{O_1} \gamma^\mu\gamma_5
\{A_\mu,O_1\} - F_{O_1}\,\textrm{Tr}\,\overline{O_1}
\gamma^\mu\gamma_5 [A_\mu,O_1],
\\
{\cal L}_{O_2} & = & \textrm{Tr}\,\overline{O_2} (i
\mathcal{D}\!\!\!\!/-m_{O_2}) O_2 \nonumber\\
&&- D_{O_2}\,\textrm{Tr}\,\overline{O_2} \gamma^\mu\gamma_5
\{A_\mu,O_2\} - F_{O_2}\,\textrm{Tr}\,\overline{O_2}
\gamma^\mu\gamma_5 [A_\mu,O_2] ,
\\
{\mathcal L}_{\Lambda_1}&=&\overline{\Lambda_1}i\partial\!\!\!/
\Lambda_1+m_{\Lambda_1}\overline{\Lambda_1}\Lambda_1,
\\
{\cal L}_Q & = &({\cal D}_\mu Q)({\cal D}^\mu\bar{Q})-m^2_Q
Q\bar{Q} ,
\\
{\cal L}_S & = & \overline{S}(i \mathcal{D}\!\!\!\!/-m_S)S
+\mathcal{G}_S \overline{S}\gamma_5 A\!\!\!/ S ,
\\
{\cal L}_T & =
&\overline{T}(i\mathcal{D}\!\!\!\!/-m_T)T+\mathcal{G}_{T}\overline{T}\gamma_5A\!\!\!/T.
\end{eqnarray}
\end{subequations}
In the above equations, $m_B$, $m_P$ etc are hadrons masses in the
chiral limit.

Keeping the flavor indices explicitly, we get
\begin{subequations}
\begin{eqnarray}
{\cal L}_{\Sigma} & = & {F_\pi^2 \over 4} \big[ ({\partial}_\mu
\Sigma^{\dagger i}_j) ({\partial}^{\mu} \Sigma^j_i) -2\mu m^i_j (
\Sigma^j_i + \Sigma^{\dagger j}_i ) \big],
\\
{\cal L}_{B} & = & \overline{B}^i_j (i
\partial\!\!\!/-m_B) B^j_i+i\overline{B}^i_j \gamma^\mu V^j_{\mu,\,k}B^k_i-i\overline{B}^i_j\gamma^\mu B^j_kV^k_{\mu,\,i} \nonumber\\
&&+ (D_B+F_B)\overline{B}^i_a \gamma_5\gamma^\mu
A^a_{\mu,\,b}B^b_i+(D_B - F_B)\overline{B}^i_a \gamma_5\gamma^\mu
B^a_b A^b_{\mu,\, i},
\\
{\cal L}_D&=&\bar{D}_{ijk}(i \partial\!\!\!/-m_D)D^{ijk}
+i\overline{D}_{ijk}\gamma^\mu D^{ija}V^{ k}_{\mu,\, a}
+i\overline{D}_{ijk}\gamma^\mu D^{iak}V^{ j}_{\mu,\, a}
+i\overline{D}_{ijk}\gamma^\mu D^{ajk}V^{ i}_{\mu,\, a}\nonumber\\
&&
+\mathcal{G}_{D} \overline{D}_{ija}\gamma_5\gamma^\mu
A_{\mu,\,b}^a D^{bij}
,\\
{\cal L}_P & = & \overline{P}^{ijk}(i \partial\!\!\!/-m_P)P_{ijk}
+i\overline{P}^{ijk}\gamma^\mu P_{ija}V^{\dagger a}_{\mu,\, k}
+i\overline{P}^{ijk}\gamma^\mu P_{iak}V^{\dagger a}_{\mu,\, j}
+i\overline{P}^{ijk}\gamma^\mu P_{ajk}V^{\dagger a}_{\mu,\,
i}\nonumber\\
&& +\mathcal{G}_{P} \overline{P}^{ija}\gamma_5\gamma^\mu
A_{\mu,\,a}^b P_{bij}
\\
{\cal L}_{O_1} & = & \overline{O_1}^i_j (i
\partial\!\!\!/-m_{O_1}) {O_1}^j_i+i\overline{O_1}^i_j \gamma^\mu V^j_{\mu,\,k}{O_1}^k_i-i\overline{O_1}^i_j\gamma^\mu {O_1}^j_kV^k_{\mu,\,i} \nonumber\\
&&+(D_{O_1}+F_{O_1})\overline{O_1}^i_a \gamma_5\gamma^\mu
A^a_{\mu,\,b}{O_1}^b_i+(D_{O_1}- F_{O_1})\overline{O_1}^i_a
\gamma_5\gamma^\mu {O_1}^a_b A_{\mu,\,i}^b
\\
{\cal L}_{O_2} & = & \overline{O_2}^i_j (i
\partial\!\!\!/-m_{O_2}) {O_2}^j_i+i\overline{O_2}^i_j \gamma^\mu V^j_{\mu,\,k}{O_2}^k_i-i\overline{O_2}^i_j\gamma^\mu {O_2}^j_kV^k_{\mu,\,i} \nonumber\\
&&+(D_{O_2}+F_{O_2})\overline{O_2}^i_a \gamma_5\gamma^\mu
A^a_{\mu,\,b}{O_2}^b_i+(D_{O_2}- F_{O_2})\overline{O_2}^i_a
\gamma_5\gamma^\mu {O_2}^a_b A_{\mu,\,i}^b
\\
{\mathcal L}_{\Lambda_1}&=&\overline{\Lambda_1}i\partial\!\!\!/
\Lambda_1+m_{\Lambda_1}\overline{\Lambda_1}\Lambda_1
\\
{\cal L}_Q & = &(\partial_\mu Q_i)(\partial^\mu \bar{Q}^i)+
\partial_\mu Q_i V^{\mu,\,i}_j \bar{Q}^j+ Q_i V^{\dag i}_{\mu,\,j}\partial^\mu \bar{Q}^j
+Q_i V^{\dag i}_{\mu,\,a} V^{\mu,\,a}_j \bar{Q}^j  -m^2_Q Q_i
\bar{Q}^i ,
\\
{\cal L}_S & = & \overline{S}^{ij}(i \partial\!\!\!/-m_S)S_{ij}
+i\overline{S}^{ij}\gamma^\mu S_{ia}V^{\dagger a}_{\mu,\, j}
+i\overline{S}^{ij}\gamma^\mu S_{aj}V^{\dagger a}_{\mu,\,
i}\nonumber\\
&& +\mathcal{G}_S \overline{S}^{ia}\gamma_5\gamma^\mu
A^b_{\mu,\,a} S_{bi}
\\
{\cal L}_T & =
&\overline{T}_i(i\partial\!\!\!/-m_T)T^i+i\overline{T}_i
V\!\!\!\!/^i_j T^j+\mathcal{G}_{T}\overline{T}_i\gamma_5\gamma^\mu
A^i_{\mu,\,j}T^j.
\end{eqnarray}
\end{subequations}

\subsection{Interaction between different matter fields}

The interaction part of the chiral Lagrangians between different
matter fields reads
\begin{subequations}
\begin{eqnarray}
 {\cal L}_{PAB} &=&\mathcal{C}_{PAB}
\big(\overline{P}\Gamma_P A\!\!\!/ B +\overline{B} \Gamma_P
A\!\!\!/ P \big),
\\
{\cal L}_{O_1AD}&=&{\cal C}_{O_1AD}\overline{O_1} A^\mu D_\mu +
h.c.
,\\
{\cal L}_{O_2AD}&=&{\cal C}_{O_2AD}\overline{O_2} i\gamma_5 A^\mu D_\mu + h.c.  ,\\
{\cal L}_{O_1AB}
&=&\mathcal{C}_{O_1AB}\textrm{Tr}\big(\overline{O_1}\,\gamma_5
\gamma^\mu\{A_\mu,B\}+\overline{B}\,\gamma_5
\gamma^\mu\{A_\mu,O_1\}\big) \nonumber\\
&&+\mathcal{H}_{O_1AB}\textrm{Tr}\big(\overline{O_1}\,\gamma_5
\gamma^\mu[A_\mu,B]+\overline{B}\,\gamma_5
\gamma^\mu[A_\mu,O_1]\big),
\\
{\cal L}_{O_1AP} &=&\mathcal{C}_{O_1AP} \big(\overline{O_1}
\Gamma_P A\!\!\!/ P+\overline{P}\Gamma_P A\!\!\!/ O_1 \big),
\\
{\cal L}_{O_2AB}
&=&\mathcal{C}_{O_2AB}\textrm{Tr}\big(\overline{O_2}
\gamma^\mu\{A_\mu,B\}+\overline{B}
\gamma^\mu\{A_\mu,O_2\}\big) \nonumber\\
&&+\mathcal{H}_{O_2AB}\textrm{Tr}\big(\overline{O_2}
\gamma^\mu[A_\mu,B]+\overline{B} \gamma^\mu[A_\mu,O_2]\big),
\\
{\cal L}_{O_2AP} &=& \mathcal{C}_{O_2AP} \big(\overline{O_2}
\Gamma_P\gamma_5 A\!\!\!/ P+\overline{P}\Gamma_P\gamma_5 A\!\!\!/
O_2 \big),
\\
{\cal L}_{O_2AO_1}
&=&\mathcal{C}_{O_2AO_1}\textrm{Tr}\big(\overline{O_2}
\gamma^\mu\{A_\mu,O_1\}+\overline{O_1}
\gamma^\mu\{A_\mu,O_2\}\big) \nonumber\\
&&+\mathcal{H}_{O_2AO_1}\textrm{Tr}\big(\overline{O_2}
\gamma^\mu[A_\mu,O_1]+\overline{O_1} \gamma^\mu[A_\mu,O_2]\big),
\\
{\cal L}_{\Lambda_1AB}
&=&\mathcal{C}_{\Lambda_1AB}\textrm{Tr}\big(\overline{\Lambda_1}\Gamma_{\Lambda_1}A\!\!\!/B
+\overline{B}\Gamma_{\Lambda_1}A\!\!\!/\Lambda_1\big),
\\
{\cal L}_{\Lambda_1AO_1}
&=&\mathcal{C}_{\Lambda_1AO_1}\textrm{Tr}\big(\overline{\Lambda_1}\Gamma_{\Lambda_1}A\!\!\!/O_1
+\overline{O_1}\Gamma_{\Lambda_1}A\!\!\!/\Lambda_1\big),
\\
{\cal L}_{\Lambda_1AO_2}
&=&\mathcal{C}_{\Lambda_1AO_2}\textrm{Tr}\big(\overline{\Lambda_1}\Gamma_{\Lambda_1}\gamma_5A\!\!\!/O_2
+\overline{O_2}\Gamma_{\Lambda_1}\gamma_5A\!\!\!/\Lambda_1\big),
\\
{\cal L}_{SQB} &=&\mathcal{C}_{SQB} \overline{S}\Gamma_{S}\bar{Q}
B +h.c ,
\\
{\cal L}_{SQP}
&=&\mathcal{C}_{SQP}\overline{S}\Gamma_{SP}\bar{Q}P+h.c,
\\
{\cal L}_{SQO_1}
&=&\mathcal{C}_{SQO_1}\overline{S}\Gamma_{S}\bar{Q}O_1+h.c,
\\
{\cal L}_{SQO_2}
&=&\mathcal{C}_{SQO_2}\overline{S}\Gamma_{S}\gamma_5\bar{Q}O_2+h.c,
\\
{\cal L}_{TQB} &=&\mathcal{C}_{TQB}\overline{T}\Gamma_T B\bar{Q}
+h.c ,
\\
{\cal L}_{TQO_1}
&=&\mathcal{C}_{TQO_1}\overline{T}\Gamma_{T}O_1\bar{Q}+h.c,
\\
{\cal L}_{TQO_2}
&=&\mathcal{C}_{TQO_2}\overline{T}\Gamma_{T}\gamma_5
O_2\bar{Q}+h.c,
\\
{\cal L}_{TQ\Lambda_1}
&=&\mathcal{C}_{TQ\Lambda_1}\overline{T}\Gamma_{T}
\bar{Q}\Lambda_1+h.c,
\\
{\cal L}_{TAS}
&=&\mathcal{C}_{TAS}(\overline{T}\Gamma_{TS}A\!\!\!/S+\overline{S}\Gamma_{TS}A\!\!\!/T),
\end{eqnarray}
\end{subequations}
where $D^\mu$ is the Rarita-Schwinger spinor for the $\Delta$
decuplet, the subscripts $P,S,T$ are the parities of pentaquarks
(anti-decuplet, anti-sextet, triplet, respectively), and the
subscript $SP$ is the product of $S$ and $P$. $\Gamma_+ =
\gamma_5$, and $\Gamma_- = 1$.

With explicit flavor indices we have
\begin{subequations}
\begin{eqnarray}
{\cal L}_{PAB} &=&\mathcal{C}_{PAB}\overline{P}^{ijk}\Gamma_P
A\!\!\!/^a_i B^b_j\epsilon_{abk} +h.c. ,
\\
{\cal L}_{O_1AD} &=&{\cal C}_{O_1AD}\overline{O_1}^a_i  A^b_{\mu
j} D^{\mu ijk} \epsilon_{abk} +h.c.
,\\
{\cal L}_{O_2AD} &=&{\cal C}_{O_2AD}\overline{O_2}^a_i i\gamma_5
A^b_{\mu j} D^{\mu ijk}
\epsilon_{abk} +h.c. ,\\
{\cal L}_{O_1AB}
&=&(\mathcal{C}_{O_1AB}+\mathcal{H}_{O_1AB})\overline{O_1}^i_a\gamma_5
\gamma^\mu A^a_{\mu,\,b}B^b_i \nonumber\\
&&+(\mathcal{C}_{O_1AB}-\mathcal{H}_{O_1AB})\overline{O_1}^i_a\gamma_5
\gamma^\mu B^a_b A^b_{\mu,\,i}+h.c.,
\\
{\cal L}_{O_1AP} &=&\mathcal{C}_{O_1AP} \overline{O_1}^i_a
\Gamma_P A\!\!\!/^j_b P_{ijk}\epsilon^{abk}+h.c.,
\\
{\cal L}_{O_2AB}
&=&(\mathcal{C}_{O_2AB}+\mathcal{H}_{O_2AB})\overline{O_2}^i_a
\gamma^\mu A^a_{\mu,\,b}B^b_i \nonumber\\
&&+(\mathcal{C}_{O_2AB}-\mathcal{H}_{O_2AB})\overline{O_2}^i_a
\gamma^\mu B^a_b A^b_{\mu,\,i}+h.c.,
\\
{\cal L}_{O_2AP} &=&\mathcal{C}_{O_2AP} \overline{O_2}^i_a
\Gamma_P\gamma_5 A\!\!\!/^j_b P_{ijk}\epsilon^{abk}+h.c.,
\\
{\cal L}_{O_2AO_1}
&=&(\mathcal{C}_{O_2AO_1}+\mathcal{H}_{O_2AO_1})\overline{O_2}^i_a
\gamma^\mu A^a_{\mu,\,b}{O_1}^b_i \nonumber\\
&&+(\mathcal{C}_{O_2AO_1}-\mathcal{H}_{O_2AO_1})\overline{O_2}^i_a
\gamma^\mu {O_1}^a_b A^b_{\mu,\,i}+h.c,
\\
{\cal L}_{\Lambda_1AB}
&=&\mathcal{C}_{\Lambda_1AB}\overline{\Lambda_1}\Gamma_{\Lambda_1}A\!\!\!/^i_j
B^j_i +h.c,
\\
{\cal L}_{\Lambda_1AO_1}
&=&\mathcal{C}_{\Lambda_1AO_1}\overline{\Lambda_1}\Gamma_{\Lambda_1}A\!\!\!/^i_j
{O_1}^j_i +h.c.,
\\
{\cal L}_{\Lambda_1AO_2}
&=&\mathcal{C}_{\Lambda_1AO_2}\overline{\Lambda_1}\Gamma_{\Lambda_1}\gamma_5A\!\!\!/^i_j
{O_2}^j_i+h.c.,
\\
{\cal L}_{SQB} &=&\mathcal{C}_{SQB}\overline{S}^{ij} \Gamma_{S}
\bar{Q}^a B^b_j \epsilon_{iab} +h.c ,
\\
{\cal L}_{SQP}
&=&\mathcal{C}_{SQP}\overline{S}^{ij}\Gamma_{SP}\bar{Q}^k
P_{ijk}+h.c,
\\
{\cal L}_{SQO_1} &=&\mathcal{C}_{SQO_1}\overline{S}^{ij}
\Gamma_{S}\bar{Q}^a {O_1}^b_j\epsilon_{iab}+h.c,
\\
{\cal L}_{SQO_2} &=&\mathcal{C}_{SQO_2}\overline{S}^{ij}
\Gamma_{S}\gamma_5\bar{Q}^a {O_2}^b_j\epsilon_{iab}+h.c,
\\
{\cal L}_{TQB} &=&\mathcal{C}_{TQB}\overline{T}_i\Gamma_T
\bar{Q}^j B^i_j +h.c ,
\\
{\cal L}_{TQO_1}
&=&\mathcal{C}_{TQO_1}\overline{T}_i\Gamma_{T}\bar{Q}^j
{O_1}^i_j+h.c,
\\
{\cal L}_{TQO_2}
&=&\mathcal{C}_{TQO_2}\overline{T}_i\Gamma_{T}\gamma_5\bar{Q}^j
{O_2}^i_j+h.c,
\\
{\cal L}_{TQ\Lambda_1}
&=&\mathcal{C}_{TQ\Lambda_1}\overline{T}_i\Gamma_{T}
\bar{Q}^i\Lambda_1+h.c,
\\
{\cal L}_{TAS}
&=&\mathcal{C}_{TAS}\overline{T}_i\Gamma_{TS}A\!\!\!/^a_b
S_{aj}\epsilon^{ibj}+h.c.\;.
\end{eqnarray}
\end{subequations}

\section{Mass and Magnetic Moment Relations}\label{sec3}

\subsection{Mass relations}

We include the nonzero current quark mass correction, which
induces mass splitting in the multiplet. These symmetry breaking
terms for various pentaquark multiplets are
\begin{eqnarray}\label{pp}
L_P &=& \alpha_P \,\overline{P} ( \xi m \xi + \xi^{\dagger} m
\xi^{\dagger} ) P,\\ \label{o1}
L_{O_1}&=& \alpha_{O_1}
\,\textrm{Tr}\big(d_1\overline{O_1}\{\xi m \xi + \xi^{\dagger} m
\xi^{\dagger},O_1\} + f_1 \overline{O_1}
[\xi m \xi + \xi^{\dagger} m \xi^{\dagger},O_1]\big) \nonumber\\
&& +\beta_{O_1} \, \textrm{Tr}(\overline{O_1}O_1) \textrm{Tr} ( m
\Sigma + \Sigma^{\dagger} m ),\\  \label{o2}
L_{O_2}&=&
\alpha_{O_2} \,\textrm{Tr}\big(d_2\overline{O_2}\{\xi m \xi +
\xi^{\dagger} m \xi^{\dagger},O_2\} + f_2 \overline{O_2}
[\xi m \xi + \xi^{\dagger} m \xi^{\dagger},O_2]\big) \nonumber\\
&& +\beta_{O_2} \, \textrm{Tr}(\overline{O_2}O_2) \textrm{Tr} ( m
\Sigma + \Sigma^{\dagger} m ),\\ \label{ss}
L_S &=&\alpha_S
\,\overline{S} ( \xi
m \xi + \xi^{\dagger} m \xi^{\dagger} ) S.   
\end{eqnarray}

Expanding Eq. (\ref{pp}), we get the mass splittings $\Delta
m_i\equiv m_i-m_{penta}$ for pentaquark anti-decuplet $P$
\begin{subequations}
\begin{eqnarray}
\Delta m_\Theta & = & 2 \alpha_P m_s,
\\
\Delta m_{N_{10}} & = & \frac23\alpha_P \left( \hat{m} + 2 m_s
\right),
\\
\Delta m_{\Sigma_{10}} & = & \frac23 \alpha_P \left( 2 \hat{m} +
m_s \right),
\\
\Delta m_{\Xi_{10}} & = &  2~ \alpha_P \hat{m}.
\end{eqnarray}
\end{subequations}
From the above mass splittings, we can derive the following mass
relations.
\begin{subequations}
\begin{eqnarray}
m_{N_{10}}-m_{\Sigma_{10}}&=&m_\Theta-m_{N_{10}},\\
m_{\Sigma_{10}}-m_{\Xi_{10}}&=&m_{N_{10}}-m_{\Sigma_{10}}.
\end{eqnarray}
\end{subequations}
These relations have already been derived using the chiral soliton
model \cite{diak} 
and chiral Lagrangian approach \cite{lee,ko}. The equal splitting
for anti-decuplet pentaquark was also discussed in Ref.
\cite{carlson}.

Similarly, for the pentaquark octet $O_{2}$
\begin{subequations}
\begin{eqnarray}
\Delta m_{N_{8_2}} & = & [2\beta_{O_2} + \alpha_{O_2} (d+f)]
(2\hat{m}) + [\beta_{O_2}+\alpha_{O_2}(d-f)] (2m_s),
\\
\Delta m_{\Sigma_{8_2}}&=&(\beta_{O_2}+\alpha_{O_2}d) (4\hat{m}) +
2 \beta_{O_2}m_s,
\\
\Delta m_{\Xi_{8_2}}&=& [2\beta_{O_2} + \alpha_{O_2} (d-f)]
(2\hat{m}) + [\beta_{O_2}+\alpha_{O_2}(d+f)] (2m_s),
\\
\Delta m_{\Lambda_{8_2}}&=&(\beta_{O_2}+\frac13 \alpha_{O_2} d )
(4 \hat{m})+(\beta_{O_2}+\frac43\alpha_{O_2}d)(2m_s).
\end{eqnarray}
\end{subequations}
Hence we have the mass relation
\begin{equation}
2M_{N_8}+2M_{\Xi_8}=3M_{\Lambda_8}+M_{\Sigma_8},
\end{equation}
which was first derived in Ref. \cite{zhanga}. The pentaquark
octet $O_1$ has similar expression. The mass relations for ideally
mixed pentaquark anti-decuplet $P$ and pentaquark octet $O_1$ have
been discussed in Ref. \cite{lee}.

For the heavy pentaquark anti-sextet $S_c$ and $S_b$ we get
\begin{subequations}
\begin{eqnarray}
\Delta m_{\Xi_{5Q}}&=&2\alpha_{S_Q} \hat{m},
\\
\Delta m_{\Sigma_{5Q}}&=&\alpha_{S_Q} (\hat{m}+m_s),
\\
\Delta m_{\Theta_{5Q}}&=&2\alpha_{S_Q} m_s.
\end{eqnarray}
\end{subequations}
\begin{equation}
M_{\Xi_{5Q}}-M_{\Sigma_{5Q}}=M_{\Sigma_{5Q}}-M_{\Theta_{5Q}}\; .
\end{equation}
The heavy pentaquark mass splittings have been discussed in Refs
\cite{jaffe,lipkin,he,huang2}. Especially in the diquark model it
is very simple to derive this mass relation with the Hamiltonian
$H_s=M+n_s(m_s+\alpha)$.

\subsection{Heavy pentaquark magnetic moment relations}

As in Refs. \cite{zhanga,coleman,bijker}, we can derive the
magnetic moment relations of heavy pentaquark anti-sextet
\cite{huang2} in Jaffe and Wilczek's model. Interested readers are
referred to Refs. \cite{huang2,zhanga,huang,huang-1,huang-2} for
details. Here we list the results only.
\begin{subequations}
\begin{eqnarray}
\mu_{\Xi_{5c}^0}+\mu_{\Xi_{5c}^{--}}&=&2\mu_{\Xi_{5c}^-},\\
3\mu_{\Theta_c^0}-\mu_{\Sigma_{5c}^0}-2\mu_{\Sigma_{5c}^-}&=&\mu_{\Xi_{5c}^0}-\mu_{\Xi_{5c}^{--}},\\
\mu_{\Xi_{5b}^+}+\mu_{\Xi_{5b}^{-}}&=&2\mu_{\Xi_{5b}^0},\\
3\mu_{\Theta_b^+}-\mu_{\Sigma_{5b}^+}-2\mu_{\Sigma_{5b}^0}&=&\mu_{\Xi_{5b}^+}-\mu_{\Xi_{5b}^{-}}\;
.
\end{eqnarray}
\end{subequations}
These relations hold for both $J^P=\frac12^+$ and $J^P=\frac32^+$
anti-sextet in JW's model.

The magnetic moments of $J^P=\frac12^-$ heavy pentaquark triplet
in the diquark model are all identical because they come from
heavy anti-quark only.

\section{Possible strong decays and coupling constants}\label{sec4}

Besides the possible decays of pentaquarks into conventional
hadrons, we also consider the strong interactions and possible
transitions between pentaquark multiplets. If pentaquarks are
bound by flux tubes and have the non-planar diamond structure as
suggested in \cite{song}, then the possible transitions between
pentaquarks might get enhanced because of the special stable
structure although the decay phase space is smaller. Expanding the
interaction terms in the previous section we obtain the coupling
constants for different decay modes. We present the results up to
one pseudoscalar meson field. Since some interaction terms have
similar flavor structure, it's enough to consider the following
pieces: ${\cal L}_{\Lambda_1AO_1}$, ${\cal L}_{O_1AP}$, ${\cal
L}_{O_2AO_1}$, ${\cal L}_{SQO_1}$, ${\cal L}_{SQP}$, ${\cal
L}_{TQB}$ and ${\cal L}_{TAS}$.

\subsection{Possible strong decays of anti-decuplet pentaquark $P_{ijk}$}

$SU(3)$ symmetry forbids the anti-decuplet to decay into the
$\Delta$ decuplet and pion octet or the $\Lambda_1 \pi$. The
chiral Lagragian and couplings of the anti-decuplet with
pseudoscalar meson octet $M$ and nucleon octet $B$ can be found in
Refs. \cite{mehen,lee}.

The anti-decuplet pentaquarks $P_{ijk}$ and the octet $O_1$
pentaquarks lie close to each other \cite{jaffe}. Especially some
states are nearly degenerate and mix ideally. So the strong
interaction between these two multiplets is very important. One
example is the identification of $N(1440)$ and $N(1710)$ as
nucleon-like pentaquarks in the diquark model \cite{jaffe}. Such a
big mass splitting after the diagonalization of the mixing mass
matrix will allow the pionic transition to occur kinematically. We
collect the couplings of the pentaquark anti-decuplet with
even-parity pentaquark octet and pseudoscalar octet in Table
\ref{tab2:PAO1}.

We want to emphasize that the odd-parity pentaquark octet $O_2$
lies much lower than the anti-decuplet. Pionic decay modes $P\to
O_2 \pi$ are allowed kinematically in many channels. The coupling
constants can also be found from Table \ref{tab2:PAO1}.

Replacing the octet ${O_1}^i_j$ with corresponding $B^i_j$ in
Table \ref{tab2:PAO1}, one gets the coupling constants of
pentaquark anti-decuplet with nucleon octet and pseudoscalar meson
octet. We note that there is a sign difference in some terms
compared with those in Ref. \cite{lee}.

\subsection{Possible strong decays of light pentaquark octet $O_{1,2}$}

The couplings of light pentaquark octet $O_{1,2}$ with
pseudoscalar meson octet $M$ and nucleon octet $B$ can be found in
Ref. \cite{lee,zhanga}.

The even-parity and odd-parity octet pentaquarks can also decay
into the $\Delta$ decuplet and the pseudoscalar meson octet. Jaffe
and Wilczek pointed out that the decay mode $\Xi_5^-\to \Xi^{\ast
0} \pi^-$ observed by NA49 Collaboration may indicate the possible
existence of the even-parity octet \cite{jaffe}. Since these decay
modes can be measured in the near future, we present the couplings
of the octet pentaquarks $O_{1,2}$ with the decuplet baryon and
the pion octet in Table \ref{tab1:O1AD}.

Similarly, since the odd-parity octet $O_2$ is lower than the
even-parity octet $O_1$, pionic decay modes $O_1\to O_2\pi$ are
allowed kinematically in some channels. The couplings are
collected in \ref{tab3:O2AO1}. One can also get the coupling
constants of pentaquark octet with nucleon octet and pseudoscalar
meson octet from Table \ref{tab3:O2AO1} with special $b$ and
proper replacement \cite{lee,zhanga}.

It's straightforward to derive the coupling of pentaquark singlet
$\Lambda_1$ with pentaquark octet ${O_1}^i_j$ and pseudoscalar
meson octet $\pi^i_j$:
\begin{eqnarray}
{\cal L}_{\Lambda_1 AO_1}&=&-\frac{1}{F_\pi}{\cal C}_{\Lambda_1
AO_1}\overline{\Lambda_1}\Gamma_{\Lambda_1}(\partial\!\!\!/\pi^0\Sigma_{8,1}^0
+\partial\!\!\!/\pi^+\Sigma_{8,1}^-+\partial\!\!\!/\pi^-\Sigma_{8,1}^+\nonumber\\
&&+\partial\!\!\!/K^+\Xi_{8,1}^-+\partial\!\!\!/K^0\Xi_{8,1}^0
+\partial\!\!\!/K^-p_{8,1}+\partial\!\!\!/\bar{K^0}n_{8,1})+h.c.\;.
\end{eqnarray}

\subsection{Possible strong decays of heavy pentaquarks}

The interaction of heavy pentaquarks with the heavy vector meson
and nucleon octet has the same flavor structure as in the case of
heavy pseudoscalar mesons. It is interesting to note that the
heavy pentaquark observed by H1 Collaboration sits right on the
threshold of $\Delta$ and D meson. One may wonder whether this
resonance is affected largely by the threshold behavior. However,
in the $SU(3)$ symmetry limit the heavy pentaquark anti-sextet can
not decay into an decuplet and a heavy pseudoscalar meson. In
other words, this state can not be explained as a coupled channel
effect between $D^{\ast -} p$ and $D \Delta$ through t-channel
pion exchange. The anti-sextet will not decay into $\Lambda_1$
plus a heavy meson. Similarly, the heavy triplet will not decay
into the $\Delta$ decuplet plus a heavy meson or the anti-decuplet
plus a heavy meson.

The interaction between heavy pentaquarks,  nucleon octet $B$ and
pseudoscalar meson octet $M$ are discussed in \cite{wise,he}. In
Jaffe and Wilczek's diquark model, the odd-parity heavy pentaquark
triplet is much lower than the even-parity heavy sextet. Pionic
decays $S\to T \pi$ may happen in many channels. It will be very
interesting to explore this kind of decay process experimentally.
Now the heavy quark acts as a spectator. We collect the relevant
coupling constants in Table \ref{tab4:TAS}.

We list the couplings of the heavy pentaquark sextet with the
light pentaquark octet $O_{1,2}$ and the heavy pseudoscalar mesons
in Table \ref{tab5:SQO_1}, those of the sextet with anit-decuplet
and heavy mseon triplet in Table \ref{tab6:SQP}. The couplings of
the heavy pentaquark triplet with light pentaquark octet $O_{1,2}$
and heavy meson triplet are presented in Table \ref{tab7:TQB}. All
these processes might be forbidden by kinematics.

\subsection{Interaction of pentaquarks within the same multiplet}

For completeness, we also consider the interaction within the same
pentaquark multiplet arising from these terms: ${\cal
G}_P\bar{P}\gamma_5 A \!\!\!/P$, $
(D_{O}+F_{O})\textrm{Tr}(\overline{O} \gamma_5\gamma^\mu A_\mu O)+
(D_{O}- F_{O})
 \textrm{Tr}(\overline{O} \gamma_5\gamma^\mu O A_\mu)$, ${\cal G}_S\bar{S}\gamma_5 A
\!\!\!/S$ and ${\cal G}_T\bar{T}\gamma_5 A \!\!\!/T$. The coupling
constants for pentaquark octet $O_1$ or $O_2$ can be found in
Table \ref{tab3:O2AO1} through simple replacement. We collect
other couplings in Table \ref{tab8:PAP}-\ref{tab10:TAT}.

\subsection{Decay widths}

There are four types of interaction terms corresponding to four
kinds of Lorentz structures depending on the parities of the
pentaquarks.
\begin{eqnarray}
&a_1\bar{F_1}\gamma_5\gamma^\mu\partial_\mu M F_2\nonumber\\
&a_2\bar{F_1}\gamma^\mu\partial_\mu M F_2\nonumber\\
&a_3\bar{F_1}\gamma^5\bar{Q}F_2\nonumber\\
&a_4\bar{F_1}\bar{Q}F_2 ,
\end{eqnarray}
where $F_1$, $F_2$ denotes initial and final fermions
respectively, $M$ and $\bar{Q}$ are the pseudoscalar mesons. The
corresponding decay widths are
\begin{eqnarray}\label{Gamma:4}
\Gamma_1 &=& \frac{|\textbf{p}^*|}{8\pi m_1^2}~a_1^2
(m_1+ m_2)^2~[(m_1 - m_2)^2-m_M^2] \nonumber\\
\Gamma_2 &=& \frac{|\textbf{p}^*|}{8\pi m_1^2}~a_2^2
(m_1 - m_2)^2~[(m_1 + m_2)^2-m_M^2] \nonumber\\
\Gamma_3 &=& \frac{|\textbf{p}^*|}{8\pi m_1^2}~a_3^2
~[(m_1 - m_2)^2-m_Q^2] \nonumber\\
\Gamma_4 &=& \frac{|\textbf{p}^*|}{8\pi m_1^2}~a_4^2 ~[(m_1 +
m_2)^2-m_Q^2],
\end{eqnarray}
where $\textbf{p}^*$ is the meson momentum in the parent particle
$F_1$ rest frame.
\begin{equation}
|\textbf{p}^*|^2=\frac{1}{4m_1^2}[m_1^2-(m_2+m_M)^2][m_1^2-(m_2-m_M)^2].
\end{equation}

For example, $\Theta^+$ width reads
\begin{eqnarray}
\Gamma_{\Theta^+}&=&2\Gamma_{\Theta^+\rightarrow
K^+n}=2\Gamma_{\Theta^+\rightarrow K^0p}\nonumber\\
&=&\frac{\mathcal{C}^2_{PAB}|\textbf{p}_1|}{4\pi F_\pi^2
m_\Theta^2} (m_\Theta+ m_N)^2~[(m_\Theta- m_N)^2-m_K^2].
\end{eqnarray}

If the mass of $\Lambda_1$ is around 1405 MeV \cite{zhanga}, it
decays into $\pi\Sigma$ only. Its width is
\begin{eqnarray}
\Gamma_{\Lambda_1}&=&3\Gamma_{\Lambda_1\rightarrow
\pi^+\Sigma^-}\nonumber\\
&=&\frac{3\mathcal{C}^2_{\Lambda_1AB}|\textbf{p}_2|}{8\pi F_\pi^2
m_{\Lambda_1}^2} (m_{\Lambda_1}- m_\Sigma)^2~[(m_{\Lambda_1}+
m_\Sigma)^2-m_\pi^2].
\end{eqnarray}
Even with $|\mathcal{C}_{\Lambda_1AB}|=10|\mathcal{C}_{PAB}|$,
hence $\Gamma_{\Lambda_1}/\Gamma_{\Theta^+}\approx 43$,
$\Lambda_1$ is still not a broad resonance assuming the current
experimental upper bound of $\Theta^+$ width.

The ratio $BR[\Xi^{--}_{10}\to \Sigma^-K^-]/BR[\Xi^{--}_{10}\to
\Xi^-\pi^-]$ are different for positive and negative parity
pentaquark, which is independent of models. This property has been
proposed to determine the parity of pentaquark anti-decuplet in
Ref. \cite{mehen}. If we assume
$\Gamma\left(\Xi^{--}_{10}\to\Xi^-_{8,2}\pi^-\right) $ is
significantly smaller than $\Gamma\left(\Xi^{--}_{10}\to
\Sigma^-K^-\right)$ and $\Gamma\left(\Xi^{--}_{10}\to
\Xi^-\pi^-\right)$ because of phase space suppression, then the
ratio $\Gamma_{\Xi^{--}_{10}}/\Gamma_{\Theta^+}$ is about $4.1$
for positive parity and $2.0$ for negative parity with
$m_{\Xi^{--}}=1862$ MeV. If their widths can be measured
accurately, the parity of the anti-decuplet can be determined
\cite{mehen}.

\section{Summary and Discussions}

We have constructed the chiral Lagrangian involving six $SU(3)_f$
pentaquark multiplets. In the framework of Jaffe and Wilczek's
diquark model, these pentaquark muptiplets include one even-parity
anti-decuplet, one even-parity octet, one odd-parity octet, one
odd-parity singlet, one even-parity heavy anti-sextet, and one
heavy triplet. However our discussion relies on the $SU(3)$
symmetry only. Therefore the results are general and not limited
to this particular model.

After taking into account of the symmetry breaking correction from
the non-zero quark mass, we have derived the Gell-Mann$-$Okubo
mass relations for different pentaquark multiplets. Similarly, we
have also derived the Coleman-Glashow relations for heavy
pentaquark magnetic moments. We have discussed the couplings of
pentaquarks with other pentaquarks and pseudoscalar mesons. We
have also investigated the possible decays of pentaquarks into the
$\Delta$ decuplet and pseudoscalar mesons.

If symmetry and kinematics allow, the most efficient decay
mechanism of pentaquarks is for the four quarks and one anti-quark
to regroup with each other into a three-quark baryon and a meson.
This is in contrast to the $^3P_0$ decay models for the ordinary
hadrons. This regrouping is coined as the "fall-apart" mechanism,
which leads to selection rules in the octet pentaquark decays.
This "fall-apart" decay mechanism can be taken care of in the
chiral Lagrangian formalism through keeping the flavor indices
explicitly \cite{lee,zhanga}. The couplings of two octet baryons
with a pseudoscalar mesons with the general F/D flavor structure
is presented in Table \ref{tab3:O2AO1}. It is pointed out that the
"fall-apart" mechanism requires $b={1\over 3}$ for the even-parity
pentaquark octet decays into nucleon octet and pseudoscalar meson
\cite{lee}. In contrast, this mechanism requires $b=-1$ for the
odd-parity pentaquark octet decays into nucleon octet and
pseudoscalar meson \cite{zhanga}.

We collect all the possible decay modes of $\Theta^+$,
$\Xi^{--}_{10}$ and $\Theta^0_c$ in Table \ref{tab11} with
corresponding coupling constants in JW's model. We find that
$\Xi^{--}_{10}$ can also decay into $\Xi^-_{8,2}$ via the emission
of a $\pi^-$. The heavy pentaquark $\Theta^0_c$ has four decay
channels, $D^-p$, $\bar{D}^0 n$, $D^{*-}p$ and $\bar{D}^{*0} n$.
The decay modes and couplings of the other exotic anti-decuplet
members and anti-sextet are also included in the table. Using the
the mass of $\Theta^0_c$ from H1 experiment as a constraint, we
have updated our old mass estimate of heavy pentaquarks in
\cite{huang2} and use the new values to analyze the possible decay
modes in Table \ref{tab11}. Hopefully our present study may help
the future experimental discovery of those missing pentaquarks.

S.L.Z. thanks Prof W.-Y. P. Hwang and COSPA center at National
Taiwan University for the hospitality. This project was supported
by the National Natural Science Foundation of China under Grant
10375003, Ministry of Education of China, FANEDD and SRF for ROCS,
SEM.


\begin{table}[h]
\begin{center}
\begin{tabular}{cc|cc|cc|cc}\hline
\multicolumn{2}{c}{$\Theta^+$} & \multicolumn{2}{c}{$N_{10}^+$} &
\multicolumn{2}{c}{$N_{10}^0$} &
\multicolumn{2}{c}{$\Sigma_{10}^+$}
\\ \hline
$K^+ n_{8,1}$ & 1 & $\pi^+ n_{8,1}$ & $-\frac{1}{\sqrt3}$ & $\pi^0
n_{8,1}$ &
$\frac{1}{\sqrt6}$ & $\pi^+ \Lambda_{8,1}$ & $\frac{1}{\sqrt2}$ \\
$K^0 p_{8,1}$ & $-1$ & $\pi^0 p_{8,1}$ & $-\frac{1}{\sqrt6}$ &
$\pi^- p_{8,1}$ & $-\frac{1}{\sqrt3}$
& $\pi^+ \Sigma^0_{8,1}$ & $-\frac{1}{\sqrt6}$ \\
& & $\eta_0 p_{8,1}$ & $\frac{1}{\sqrt2}$ & $\eta_0 n_{8,1}$ &
$\frac{1}{\sqrt2}$ &
$\pi^0 \Sigma^+_{8,1}$ & $\frac{1}{\sqrt6}$ \\
& & $K^+ \Lambda_{8,1}$ & $-\frac{1}{\sqrt2}$ & $K^+
\Sigma^-_{8,1}$ &
$\frac{1}{\sqrt3}$ & $\eta_0 \Sigma^+_{8,1}$ & $-\frac{1}{\sqrt2}$ \\
& & $K^+ \Sigma^0_{8,1}$ & $\frac{1}{\sqrt6}$ & $K^0
\Lambda_{8,1}$ &
$-\frac{1}{\sqrt2}$ & $K^+ \Xi^0_{8,1}$ & $-\frac{1}{\sqrt3}$ \\
& & $K^0 \Sigma^+_{8,1}$ & $\frac{1}{\sqrt3}$ & $K^0
\Sigma^0_{8,1}$ &
$-\frac{1}{\sqrt6}$ & $\bar{K}^0 p_{8,1} $ & $\frac{1}{\sqrt3}$ \\
\hline \multicolumn{2}{c}{$\Sigma_{10}^0$} &
\multicolumn{2}{c}{$\Sigma_{10}^-$} &
\multicolumn{2}{c}{$\Xi^+_{10}$} &
\multicolumn{2}{c}{$\Xi^0_{10}$} \\ \hline $\pi^+ \Sigma^-_{8,1}$
& $-\frac{1}{\sqrt6}$ & $\pi^0 \Sigma^-_{8,1}$ &
$\frac{1}{\sqrt6}$ & $\pi^+\Xi^0_{8,1}$ & $1$ & $\pi^+ \Xi^-_{8,1}$ & $-\frac{1}{\sqrt3}$ \\
$\pi^0 \Lambda_{8,1}$ & $-\frac{1}{\sqrt2}$ & $\pi^-
\Lambda_{8,1}$ & $-\frac{1}{\sqrt2}$
& $\bar{K}^0 \Sigma^+_{8,1}$ & $-1$ & $\pi^0 \Xi^0_{8,1}$ & $-\sqrt{\frac{2}{3}}$ \\
$\pi^- \Sigma^+_{8,1}$ & $\frac{1}{\sqrt6}$ & $\pi^-
\Sigma^0_{8,1}$ &
$-\frac{1}{\sqrt6}$ & & & $\bar{K}^0 \Sigma^0_{8,1}$ & $\sqrt{\frac{2}{3}}$ \\
$\eta_0\Sigma^0_{8,1}$ & $\frac{1}{\sqrt2}$ & $\eta_0
\Sigma^-_{8,1}$ & $\frac{1}{\sqrt2}$ &
& & $K^-\Sigma^+_{8,1}$ & $\frac{1}{\sqrt3}$ \\
$K^+ \Xi^-_{8,1}$ & $\frac{1}{\sqrt6}$ & $K^0 \Xi^-_{8,1}$ & $\frac{1}{\sqrt3}$ & & & & \\
${K}^0 \Xi^0_{8,1}$ & $-\frac{1}{\sqrt6}$ & $K^- n_{8,1}$ & $-\frac{1}{\sqrt3}$ & & & & \\
$\bar{K}^0 n_{8,1}$ & $\frac{1}{\sqrt6}$ & & & & & & \\
${K}^- p_{8,1}$ & $-\frac{1}{\sqrt6}$ & & & & & & \\ \hline
\multicolumn{2}{c}{$\Xi^-_{10}$} &
\multicolumn{2}{c}{$\Xi^{--}_{10}$} & \multicolumn{2}{c}{} &
\multicolumn{2}{c}{} \\ \hline
$\pi^0 \Xi^-_{8,1}$ & $\sqrt{\frac{2}{3}}$ & $\pi^- \Xi^-_{8,1}$ & $1$ & & & & \\
$\pi^- \Xi^0_{8,1}$ & $-\frac{1}{\sqrt3}$ & $K^- \Sigma^-_{8,1}$ & $-1$ & & & & \\
$\bar{K}^0 \Sigma^-_{8,1}$ & $\frac{1}{\sqrt3}$ & & & & & & \\
${K}^- \Sigma^0_{8,1}$ & $-\sqrt{\frac{2}{3}}$ & & & & & & \\
\hline
\end{tabular}
\end{center}
\caption{Couplings of the pentaquark anti-decuplet $P_{ijk}$ with
the pentaquark octet ${O_1}^i_j$ and pseudoscalar meson octet
$\pi^i_j$. The universal coupling constant $-\frac{1}{F_\pi}{\cal
C}_{O_1AP}$ is omitted.} \label{tab2:PAO1}
\end{table}

\begin{table}[h]
\begin{center}
\begin{tabular}{cc|cc|cc|cc}\hline
\multicolumn{2}{c}{$\Xi_{8,1}^-$} &
\multicolumn{2}{c}{$\Xi_{8,1}^0$} & \multicolumn{2}{c}{$p_{8,1}$}
& \multicolumn{2}{c}{$n_{8,1}$}
\\ \hline
$\pi^- \Xi^0$ & $\frac{1}{\sqrt3}$ & $\pi^+ \Xi^-$ &
$-\frac{1}{\sqrt3}$ & $\pi^0 \Delta^+$ & $\sqrt{\frac23}$ &
$\pi^0 \Delta^0 $ & $\sqrt{\frac23}$ \\
$\pi^0 \Xi^-$ & $-\frac{1}{\sqrt6}$ & $\pi^0 \Xi^0$ &
$-\frac{1}{\sqrt6}$ & $\pi^+ \Delta^0$ & $\frac{1}{\sqrt3}$
& $\pi^- \Delta^+$ & $-\frac{1}{\sqrt3}$ \\
$\eta_0 \Xi^-$ & $\frac{1}{\sqrt2}$ & $\eta_0 \Xi^0$ &
$-\frac{1}{\sqrt2}$ & $\pi^- \Delta^{++}$ & $-1$ &
$\pi^+ \Delta^-$ & $1$ \\
$\bar{K}^0 \Sigma^-$ & $-\frac{1}{\sqrt3}$ & $K^- \Sigma^+$ &
$\frac{1}{\sqrt3}$ & $K^+ \Sigma^0$ &
$\frac{1}{\sqrt6}$ & $K^0 \Sigma^0$ & $-\frac{1}{\sqrt6}$ \\
$K^- \Sigma^0$ & $-\frac{1}{\sqrt6}$ & $\bar{K}^0 \Sigma^0$ &
$\frac{1}{\sqrt6}$ & $K^0 \Sigma^+$ &
$-\frac{1}{\sqrt3}$ & $K^+ \Sigma^-$ & $\frac{1}{\sqrt3}$ \\
$K^0 \Omega^-$ & $1$ & $K^+\Omega^-$ & $-1$ &&&& \\
\hline \multicolumn{2}{c}{$\Sigma_{8,1}^0$} &
\multicolumn{2}{c}{$\Sigma_{8,1}^+$} &
\multicolumn{2}{c}{$\Sigma_{8,1}^-$} &
\multicolumn{2}{c}{$\Lambda_{8,1}$}
\\ \hline
$\pi^+ \Sigma^-$ & $\frac{1}{\sqrt6}$ & $\pi^+ \Sigma^0$ &
$-\frac{1}{\sqrt6}$ & $\pi^-\Sigma^0$ & $\frac{1}{\sqrt6}$ & $\pi^+ \Sigma^-$ & $-\frac{1}{\sqrt2}$ \\
$\pi^- \Sigma^+$ & $\frac{1}{\sqrt6}$ & $\pi^0 \Sigma^+$ &
$-\frac{1}{\sqrt6}$
& $\pi^0 \Sigma^-$ & $-\frac{1}{\sqrt6}$ & $\pi^0 \Sigma^0$ & $-\frac{1}{\sqrt2}$ \\
$\eta_0 \Sigma^0 $ & $\frac{1}{\sqrt2}$ & $\eta_0 \Sigma^+$ &
$-\frac{1}{\sqrt2}$ & $\eta_0 \Sigma^-$ & $\frac{1}{\sqrt2}$ & $\pi^- \Sigma^+$ & $\frac{1}{\sqrt2}$ \\
$K^0\Xi^0$ & $\frac{1}{\sqrt6}$ & $K^+\Xi^0 $ &
$-\frac{1}{\sqrt3}$ & $K^-\Delta^0 $
& $-\frac{1}{\sqrt3}$ & $K^0\Xi^0 $ & $\frac{1}{\sqrt2}$ \\
$K^- \Delta^+$ & $-\sqrt{\frac23}$ & $K^- \Delta^{++}$ & $1$ &
$K^0 \Xi^-$ & $\frac{1}{\sqrt3}$ & $ K^+ \Xi^-$ &
$-\frac{1}{\sqrt2}$ \\
$\bar{K}^0 \Delta^0$ & $-\sqrt{\frac23}$ & $\bar{K}^0 \Delta^+$ &
$\frac{1}{\sqrt3}$ & $\bar{K}^0\Delta^- $ &$-1$ &&\\
$K^+ \Xi^-$ & $\frac{1}{\sqrt6}$ & & & & & &  \\
\hline
\end{tabular}
\end{center}
\caption{Couplings of the pentaquark octet ${O_{1}}^i_j$ with the
baryon decuplet $D^{ijk}$ and pseudoscalar meson octet $\pi^i_j$.
The universal coupling constant $-\frac{1}{F_\pi}{\cal C}_{O_1AD}$
is omitted. Except the universal coupling constant, the couplings
of $O_2$ is the same. } \label{tab1:O1AD}
\end{table}

\begin{table}[h]
\begin{center}
\begin{tabular}{cc|cc|cc|cc}\hline
\multicolumn{2}{c}{$\Xi_{8,1}^-$} &
\multicolumn{2}{c}{$\Xi_{8,1}^0$} & \multicolumn{2}{c}{$p_{8,1}$}
& \multicolumn{2}{c}{$n_{8,1}$}
\\ \hline
$\pi^- \Xi_{8,2}^0$ & $1-b$ & $\pi^+ \Xi_{8,2}^-$ & $1-b$ & $\pi^0
p_{8,2}$ & $\frac{1}{\sqrt2}(1+b)$ &
$\pi^0 n_{8,2} $ & $-\frac{1}{\sqrt2}(1+b)$ \\
$\pi^0 \Xi_{8,2}^-$ & $\frac{1}{\sqrt2}(1-b)$ & $\pi^0
\Xi_{8,2}^0$ & $-\frac{1}{\sqrt{2}}(1-b)$ & $\pi^+ n_{8,2}$ &
$1+b$
& $\pi^- p_{8,2}$ & $1+b$ \\
$\eta_0 \Xi_{8,2}^-$ & $-\frac{1}{\sqrt6}(3b+1)$ & $\eta_0
\Xi_{8,2}^0$ & $-\frac{1}{\sqrt6}(3b+1)$ & $\eta_0 p_{8,2}$ &
$\frac{1}{\sqrt6}(3b-1)$ &
$\eta_0 n_{8,2}$ & $\frac{1}{\sqrt6}(3b-1)$ \\
$\bar{K}^0 \Sigma_{8,2}^-$ & $1+b$ & $K^- \Sigma_{8,2}^+$ & $1+b$
& $K^+ \Sigma_{8,2}^0$ &
$\frac{1}{\sqrt2}(1-b)$ & $K^0 \Sigma_{8,2}^0$ & $-\frac{1}{\sqrt2}(1-b)$ \\
$K^- \Sigma_{8,2}^0$ & $\frac{1}{\sqrt2}(1+b)$ & $\bar{K}^0
\Sigma_{8,2}^0$ & $-\frac{1}{\sqrt2}(1+b)$ & $K^0 \Sigma_{8,2}^+$
&
$1-b$ & $K^+ \Sigma_{8,2}^-$ & $1-b$ \\
$K^- \Lambda_{8,2}$ & $\frac{1}{\sqrt6}(3b-1)$ & $\bar{K}^0
\Lambda_{8,2}$ & $\frac{1}{\sqrt6}(3b-1)$ & $K^+ \Lambda_{8,2}$ &
$-\frac{1}{\sqrt6}(3b+1)$ & $K^0 \Lambda_{8,2} $ & $-\frac{1}{\sqrt6}(3b+1)$ \\
\hline \multicolumn{2}{c}{$\Sigma_{8,1}^0$} &
\multicolumn{2}{c}{$\Sigma_{8,1}^+$} &
\multicolumn{2}{c}{$\Sigma_{8,1}^-$} &
\multicolumn{2}{c}{$\Lambda_{8,1}$}
\\ \hline
$\pi^+ \Sigma_{8,2}^-$ & $\sqrt2 b$ & $\pi^+ \Sigma_{8,2}^0$ &
$-\sqrt2 b$ & $\pi^-\Sigma_{8,2}^0$ & $\sqrt{2} b$ & $\pi^- \Sigma_{8,2}^+$ & $\sqrt{\frac{2}{3}}$ \\
$\pi^- \Sigma_{8,2}^+$ & $-\sqrt{2} b$ & $\pi^0 \Sigma_{8,2}^+$ &
$\sqrt{2} b$
& $\pi^0 \Sigma_{8,2}^-$ & $-\sqrt2 b$ & $\pi^+ \Sigma_{8,2}^-$ & $\sqrt{\frac{2}{3}}$ \\
$\pi^0 \Lambda_{8,2}$ & $\sqrt{\frac{2}{3}}$ & $\eta_0
\Sigma_{8,2}^+$ &
$\sqrt{\frac{2}{3}}$ & $\eta_0 \Sigma_{8,2}^-$ & $\sqrt{\frac{2}{3}}$ & $\pi^0 \Sigma_{8,2}^0$ & $\sqrt{\frac{2}{3}}$ \\
$\eta_0\Sigma_{8,2}^0$ & $\sqrt{\frac{2}{3}}$ & $\pi^+
\Lambda_{8,2}$ & $\sqrt{\frac{2}{3}}$ & $\pi^- \Lambda_{8,2}$
& $\sqrt{\frac{2}{3}}$ & $\eta_0 \Lambda_{8,2}$ & $-\sqrt{\frac{2}{3}}$ \\
$K^- p_{8,2}$ & $\frac{1}{\sqrt2}(1-b)$ & $K^+ \Xi_{8,2}^0$ & $1+b$ & $K^0 \Xi_{8,2}^-$ & $1+b$ & $ K^+ \Xi_{8,2}^-$ & $\frac{1}{\sqrt6}(3b-1)$ \\
$\bar{K}^0 n_{8,2}$ & $-\frac{1}{\sqrt2}(1-b)$ & $\bar{K}^0 p_{8,2}$ & $1-b$ & $K^- n_{8,2}$ &$1-b$ & $K^0 \Xi_{8,2}^0$ & $\frac{1}{\sqrt6}(3b-1)$ \\
$K^+ \Xi_{8,2}^-$ & $\frac{1}{\sqrt2}(1+b)$ & & & & & $K^- p_{8,2}$ & $-\frac{1}{\sqrt6}(3b+1)$ \\
${K}^0 \Xi_{8,2}^0$ & $-\frac{1}{\sqrt2}(1+b)$ & & & & &
$\bar{K}^0 n_{8,2}$ &
$-\frac{1}{\sqrt6}(3b+1)$ \\
\hline
\end{tabular}
\end{center}
\caption{Couplings of the pentaquark octet ${O_1}^i_j$ with the
pentaquark octet ${O_2}^i_j$ and pseudoscalar meson octet
$\pi^i_j$. The universal coupling constant $-\frac{1}{F_\pi}{\cal
C}_{O_2AO_1}$ is omitted. The constant $b={\cal H}_{O_2AO_1}/{\cal
C}_{O_2AO_1}$.} \label{tab3:O2AO1}
\end{table}

\begin{table}[h]
\begin{center}
\begin{tabular}{cc|cc|cc}\hline
\multicolumn{2}{c}{$\Xi_{5c}^{--}(\Xi_{5b}^{-})$} &
\multicolumn{2}{c}{$\Xi_{5c}^-(\Xi_{5b}^0)$} &
\multicolumn{2}{c}{$\Xi_{5c}^0(\Xi_{5b}^+)$}
\\ \hline
$K^- \Sigma_{5c}^{\prime-}(\Sigma_{5b}^{\prime0})$ & $1$ & $K^-
\Sigma_{5c}^{\prime0}(\Sigma_{5b}^{\prime+})$ & $\frac{1}{\sqrt2}$
&$\pi^+
\Xi_{5c}^{\prime-}(\Xi_{5b}^{\prime0})$ & $1$\\
 $\pi^- \Xi_{5c}^{\prime-}(\Xi_{5b}^{\prime0})$ & $-1$ &
$\bar{K^0}\Sigma_{5c}^{\prime-}(\Sigma_{5b}^{\prime0})$ &
$-\frac{1}{\sqrt{2}}$ &
$\bar{K^0}\Sigma_{5c}^{\prime0}(\Sigma_{5b}^{\prime+})$&$-1$\\
 & & $\pi^0 \Xi_{5c}^{\prime-}(\Xi_{5b}^{\prime0})$ & $-1$ & & \\
\hline \multicolumn{2}{c}{$\Sigma_{5c}^-(\Sigma_{5b}^0)$} &
\multicolumn{2}{c}{$\Sigma_{5c}^0(\Sigma_{5b}^+)$}&
\multicolumn{2}{c}{$\Theta_{5c}^0(\Theta_{5b}^+)$}
\\ \hline
$\pi^-\Sigma_{5c}^{\prime0}(\Sigma_{5b}^{\prime+})$ &
$\frac{1}{\sqrt2} $ &
$\pi^0\Sigma_{5c}^{\prime0}(\Sigma_{5b}^{\prime+})$ &
$\frac{1}{2}$ & $K^0\Sigma_{5c}^{\prime0}(\Sigma_{5b}^{\prime+})$ & $1$  \\
$\pi^0\Sigma_{5c}^{\prime-}(\Sigma_{5b}^{\prime0})$ &
$-\frac{1}{2}$ &
$\eta_0\Sigma_{5c}^{\prime0}(\Sigma_{5b}^{\prime+})$ &
$-\frac{\sqrt3}{2}$
& $K^+ \Sigma_{5c}^{\prime-}(\Sigma_{5b}^{\prime0})$ & $-1$ \\
$\eta_0\Sigma_{5c}^{\prime-}(\Sigma_{5b}^{\prime0})$ &
$-\frac{\sqrt3}{2}$ &
$\pi^+\Sigma_{5c}^{\prime-}(\Sigma_{5b}^{\prime0})$ &
$\frac{1}{\sqrt2}$ & &  \\
$K^0 \Xi_{5c}^{\prime-}(\Xi_{5b}^{\prime0})$&$-\frac{1}{\sqrt2}$ &
$K^+ \Xi_{5c}^{\prime-}(\Xi_{5b}^{\prime0})$ &
$-\frac{1}{\sqrt2}$ && \\
\hline
\end{tabular}
\end{center}
\caption{Couplings of the heavy pentaquark anti-sextet $S_{ij}$
with the heavy pentaquark triplet $T^i$ and pseudoscalar meson
octet $\pi^i_j$. The universal coupling constant
$-\frac{1}{F_\pi}{\cal C}_{TAS}$ is omitted.} \label{tab4:TAS}
\end{table}

\begin{table}[h]
\begin{center}
\begin{tabular}{cc|cc|cc}\hline
\multicolumn{2}{c}{$\Xi_{5c}^{--}(\Xi_{5b}^{-})$} &
\multicolumn{2}{c}{$\Xi_{5c}^-(\Xi_{5b}^0)$} &
\multicolumn{2}{c}{$\Xi_{5c}^0(\Xi_{5b}^+)$}
\\ \hline
$D^-(B^0) \Xi^-_{8,1}$ & $1$ & $\bar{D^0}(B^+) \Xi^-_{8,1}$ &
$\frac{1}{\sqrt2}$ & $\bar{D^0}(B^+) \Xi^0_{8,1}$ & $-1$ \\
$D_s^-(B_s^0) \Sigma^-_{8,1}$ & $-1$ & $D^-(B^0) \Xi^0_{8,1}$ &
$-\frac{1}{\sqrt{2}}$ & $D_s^-(B_s^0) \Sigma^+_{8,1}$ &
$1$\\
 & & $D_s^-(B_s^0) \Xi^0_{8,1}$ & $-1$ & & \\
\hline \multicolumn{2}{c}{$\Sigma_{5c}^-(\Sigma_{5b}^0)$} &
\multicolumn{2}{c}{$\Sigma_{5c}^0(\Sigma_{5b}^+)$}&
\multicolumn{2}{c}{$\Theta_{5c}^0(\Theta_{5b}^+)$}
\\ \hline
$\bar{D^0}(B^+) \Sigma^-_{8,1}$ & $\frac{1}{\sqrt2} $ &
$\bar{D^0}(B^+) \Sigma^0_{8,1}$ &
$\frac{1}{2}$ & $\bar{D^0}(B^+) n_{8,1}$ & $1$  \\
$D^-(B^0)\Sigma^0_{8,1}$ & $-\frac{1}{2}$ &
$\bar{D^0}(B^+)\Lambda_{8,1}$ & $-\frac{\sqrt3}{2}$
& $D^-(B^0)p_{8,1}$ & $-1$ \\
$D^-(B^0)\Lambda_{8,1}$ & $-\frac{\sqrt3}{2}$ &
$D^-(B^0)\Sigma^+_{8,1}$ &
$\frac{1}{\sqrt2}$ & &  \\
$D_s^-(B_s^0)n_{8,1}$&$-\frac{1}{\sqrt2}$ & $D_s^-(B_s^0)p_{8,1}$
&
$-\frac{1}{\sqrt2}$ && \\
\hline
\end{tabular}
\end{center}
\caption{Couplings of the heavy pentaquark anti-sextet $S_{ij}$
with the light pentaquark octet ${O_1}^i_j$ and heavy flavor
pseudoscalar meson triplet $\bar{Q}^i$. The universal coupling
constant ${\cal C}_{SQ O_1}$ is omitted.} \label{tab5:SQO_1}
\end{table}

\begin{table}[h]
\begin{center}
\begin{tabular}{cc|cc|cc}\hline
\multicolumn{2}{c}{$\Xi_{5c}^{--}(\Xi_{5b}^{-})$} &
\multicolumn{2}{c}{$\Xi_{5c}^-(\Xi_{5b}^0)$} &
\multicolumn{2}{c}{$\Xi_{5c}^0(\Xi_{5b}^+)$}
\\ \hline
$\bar{D^0}(B^+) \Xi_{10}^{--}$ & $1$ & $\bar{D^0}(B^+) \Xi_{10}^-$
&
$\sqrt{\frac{2}{3}}$ & $\bar{D^0}(B^+) \Xi_{10}^0$ & $\frac{1}{\sqrt3}$ \\
$D^-(B^0) \Xi_{10}^-$ & $-\frac{1}{\sqrt3}$ & $D^-(B^0)
\Xi_{10}^0$ & $-\sqrt{\frac{2}{3}}$ & $D^-(B^0) \Xi_{10}^+$ &
$-1$\\
 $D_s^-(B_s^0) \Sigma_{10}^-$& $\frac{1}{\sqrt3}$ & $D_s^-(B_s^0) \Sigma_{10}^0$ & $\frac{1}{\sqrt3}$ &$D_s^-(B_s^0)
 \Sigma_{10}^+$ & $\frac{1}{\sqrt3}$\\
\hline \multicolumn{2}{c}{$\Sigma_{5c}^-(\Sigma_{5b}^0)$} &
\multicolumn{2}{c}{$\Sigma_{5c}^0(\Sigma_{5b}^+)$}&
\multicolumn{2}{c}{$\Theta_{5c}^0(\Theta_{5b}^+)$}
\\ \hline
$\bar{D^0}(B^+) \Sigma_{10}^-$ & $\sqrt{\frac{2}{3}} $ &
$\bar{D^0}(B^+) \Sigma_{10}^0$ &
$\frac{1}{\sqrt3}$ & $\bar{D^0}(B^+)N_{10}^0$ & $\frac{1}{\sqrt3}$  \\
$D^-(B^0)\Sigma_{10}^0$ & $-\frac{1}{\sqrt3}$ &
$D^-(B^0)\Sigma_{10}^+$ & $-\sqrt{\frac{2}{3}}$
& $D^-(B^0)N_{10}^+$ & $-\frac{1}{\sqrt3}$ \\
$D_s^-(B_s^0)N_{10}^0$ & $\sqrt{\frac{2}{3}}$ &
$D_s^-(B_s^0)N_{10}^+$
&$\sqrt{\frac{2}{3}}$ & $D_s^-(B_s^0)\Theta^+$ & $1$  \\
\hline
\end{tabular}
\end{center}
\caption{Couplings of the heavy pentaquark anti-sextet $S_{ij}$
with the pentaquark anti-decuplet $P_{ijk}$ and heavy flavor
pseudoscalar meson triplet $\bar{Q}^i$. The universal coupling
constant ${\cal C}_{SQP}$ is omitted.} \label{tab6:SQP}
\end{table}
\begin{table}[h]
\begin{center}
\begin{tabular}{cc|cc|cc}\hline
\multicolumn{2}{c}{$\Sigma_{5c}^{\prime 0}(\Sigma_{5b}^{\prime
+})$} & \multicolumn{2}{c}{$\Sigma_{5c}^{\prime -}
(\Sigma_{5b}^{\prime 0})$} & \multicolumn{2}{c}{$\Xi_{5c}^{\prime
-}(\Xi_{5b}^{\prime 0})$}
\\ \hline
$\bar{D^0}(B^+)\Sigma_{8,1}^0$ & $\frac{1}{\sqrt2}$ &
$\bar{D^0}(B^+)\Sigma_{8,1}^-$ &
$1$ & $\bar{D^0}(B^+) \Xi_{8,1}^-$ & $1$ \\
$\bar{D^0}(B^+)\Lambda_{8,1}$ & $\frac{1}{\sqrt6}$ & $D^-(B^0)
\Sigma_{8,1}^0$ & $-\frac{1}{\sqrt{2}}$ & $D^-(B^0)\Xi_{8,1}^0$ &
$1$\\
$D^-(B^0)\Sigma_{8,1}^+$ & $1$ & $D^-(B^0) \Lambda_{8,1}$ & $\frac{1}{\sqrt6}$
&$D_s^-(B_s^0)\Lambda_{8,1}$ &$-\sqrt{\frac{2}{3}}$ \\
$D_s^-(B_s^0)p_{8,1}$& $1$ &$D_s^-(B_s^0) n_{8,1}$&1&& \\
\hline
\end{tabular}
\end{center}
\caption{Couplings of the heavy pentaquark triplet $T^i$ with the
pentaquark octet ${O_1}^i_j$ and heavy flavor pseudoscalar meson
triplet $\bar{Q}^i$. The universal coupling constant ${\cal
C}_{TQO_1}$ is omitted.} \label{tab7:TQB}
\end{table}

\begin{table}[h]
\begin{center}
\begin{tabular}{cc|cc|cc|cc}\hline
\multicolumn{2}{c}{$\Theta^+$} & \multicolumn{2}{c}{$N_{10}^+$} &
\multicolumn{2}{c}{$N_{10}^0$} &
\multicolumn{2}{c}{$\Sigma_{10}^+$}
\\ \hline
$K^+ N^0_{10}$ & $\frac{1}{\sqrt3}$ & $\pi^+ N^0_{10}$ &
$-\frac{1}{3}$ & $\pi^0 N^0_{10}$ &
$\frac{1}{3\sqrt2}$ & $\pi^+\Sigma^0_{10} $ & $-\frac{\sqrt2}{3}$ \\
$K^0 N^+_{10}$ & $-\frac{1}{\sqrt3}$ & $\pi^0 N^+_{10}$ &
$-\frac{1}{3\sqrt2}$ & $\pi^- N^+_{10}$ & $-\frac{1}{3}$
& $\pi^0 \Sigma^+_{10}$ & $-\frac{\sqrt2}{3}$ \\
$\eta_0 \Theta^+$ &$-\frac{2}{\sqrt6}$& $\eta_0 N^+_{10}$ &
$-\frac{1}{\sqrt6}$ & $\eta_0 N^0_{10}$ & $-\frac{1}{\sqrt6}$ &
$K^+ \Xi^0_{10}$ & $\frac{1}{3}$ \\
& & $K^+ \Sigma^0_{10}$ & $\frac{\sqrt2}{3}$ & $K^+ \Sigma^-_{10}$
&
$\frac{2}{3}$ & $K^0 \Xi^+_{10}$ & $-\frac{1}{\sqrt3}$ \\
& & $K^0 \Sigma^+_{10}$ & $-\frac{2}{3}$ & $K^-\Theta^+$ &
$\frac{1}{\sqrt3}$ & $\bar{K}^0 N^+_{10}$ & $-\frac{2}{3}$ \\
& & $\bar{K^0} \Theta^+$ & $-\frac{1}{\sqrt3}$ & $K^0
\Sigma^0_{10}$ &
$-\frac{\sqrt2}{3}$ & &  \\
\hline \multicolumn{2}{c}{$\Sigma_{10}^0$} &
\multicolumn{2}{c}{$\Sigma_{10}^-$} &
\multicolumn{2}{c}{$\Xi^+_{10}$} &
\multicolumn{2}{c}{$\Xi^0_{10}$}
\\\hline $\pi^+ \Sigma^-_{10}$ &
$-\frac{\sqrt2}{3}$ & $\pi^0 \Sigma^-_{10}$ &
$\frac{\sqrt2}{3}$ & $\pi^+\Xi^0_{10}$ & $-\frac{1}{\sqrt3}$ & $\pi^+ \Xi^-_{10}$ & $-\frac{2}{3}$ \\
$\pi^- \Sigma^+_{10}$ & $-\frac{\sqrt2}{3}$&$\pi^- \Sigma^0_{10}$
& $-\frac{\sqrt2}{3}$
& $\pi^0 \Xi^+_{10}$ & $-\frac{1}{\sqrt2}$ & $\pi^0 \Xi^0_{10}$ & $-\frac{1}{3\sqrt2}$ \\
$K^+ \Xi^-_{10}$ & $\frac{\sqrt2}{3}$ & $K^+ \Xi^{--}_{10}$ &
$\frac{1}{\sqrt3}$ & $\eta_0 \Xi^+_{10} $&$\frac{1}{\sqrt6}$& $\eta_0 \Xi^0_{10}$ & $\frac{1}{\sqrt6}$ \\
$K^- N^+_{10}$ & $\frac{\sqrt2}{3}$ & $K^0 \Xi^-_{10}$ &
$-\frac{1}{3}$ &$\bar{K}^0 \Sigma^+_{10}$
& $-\frac{1}{\sqrt3} $& $\pi^-\Xi^+_{10}$ & $-\frac{1}{\sqrt3}$ \\
$K^0 \Xi^0_{10}$ & $-\frac{\sqrt2}{3}$ & $K^- N^0_{10}$ & $\frac{2}{3}$ & & &$\bar{K}^0 \Sigma^0_{10}$ &$-\frac{\sqrt2}{3}$ \\
$\bar{K}^0 N^0_{10}$ & $-\frac{\sqrt2}{3}$ &  &  & & &$K^- \Sigma^+_{10}$ &$\frac{1}{3}$ \\
\hline \multicolumn{2}{c}{$\Xi^-_{10}$} &
\multicolumn{2}{c}{$\Xi^{--}_{10}$} & \multicolumn{2}{c}{} &
\multicolumn{2}{c}{} \\ \hline
$\pi^+ \Xi^{--}_{10}$ & $-\frac{1}{\sqrt3}$ & $\pi^0 \Xi^{--}_{10}$ & $\frac{1}{\sqrt2}$ & & & & \\
$\pi^0 \Xi^-_{10}$ & $\frac{1}{3\sqrt2}$ & $\pi^-\Xi^-_{10}$ & $-\frac{1}{\sqrt3}$ & & & & \\
$\pi^- \Xi^0_{10}$ & $-\frac{2}{3}$ &$\eta_0\Xi^{--}_{10}$ &$\frac{1}{\sqrt6}$ & & & & \\
$\eta_0 \Xi^-_{10}$&$\frac{1}{\sqrt6}$&$K^- \Sigma^-_{10}$&$\frac{1}{\sqrt3}$&&&\\
$\bar{K}^0\Sigma^-_{10}$&$-\frac{1}{3}$&&&&&&\\
${K}^- \Sigma^0_{10}$ & $\frac{\sqrt2}{3}$ & & & & & & \\
\hline
\end{tabular}
\end{center}
\caption{Couplings of the pentaquark anti-decuplet $P_{ijk}$ with
pseudoscalar meson octet $\pi^i_j$. The universal coupling
constant $-\frac{1}{F_\pi}{\cal G}_{P}$ is omitted.}
\label{tab8:PAP}
\end{table}

\begin{table}[h]
\begin{center}
\begin{tabular}{cc|cc|cc}\hline
\multicolumn{2}{c}{$\Xi_{5c}^{--}(\Xi_{5b}^{-})$} &
\multicolumn{2}{c}{$\Xi_{5c}^-(\Xi_{5b}^0)$} &
\multicolumn{2}{c}{$\Xi_{5c}^0(\Xi_{5b}^+)$}
\\ \hline
$\pi^0 \Xi_{5c}^{--}(\Xi_{5b}^{-})$ & $\frac{1}{\sqrt2}$ & $\pi^+
\Xi_{5c}^{--}(\Xi_{5b}^{-})$ & $-\frac{1}{\sqrt2}$
&$\pi^+\Xi_{5c}^-(\Xi_{5b}^0)$&$-\frac{1}{\sqrt2}$\\
 $\pi^- \Xi_{5c}^{-}(\Xi_{5b}^{0})$ & $-\frac{1}{\sqrt2}$ &
$\pi^-\Xi_{5c}^0(\Xi_{5b}^+)$ & $-\frac{1}{\sqrt{2}}$ & $\pi^0
\Xi_{5c}^0(\Xi_{5b}^+)$ & $-\frac{1}{\sqrt2}$\\
$\eta_0 \Xi_{5c}^{--}(\Xi_{5b}^{-})$ & $\frac{1}{\sqrt6}$
& $\eta_0 \Xi_{5c}^-(\Xi_{5b}^0)$ & $\frac{1}{\sqrt6}$ &$\eta_0 \Xi_{5c}^0(\Xi_{5b}^+)$ &$\frac{1}{\sqrt6}$ \\
$K^- \Sigma^-_{5c}(\Sigma^0_{5b})$&$\frac{1}{\sqrt2}$&$\bar{K}^0
\Sigma_{5c}^-(\Sigma_{5b}^0)$&$-\frac12$& $\bar{K}^0 \Sigma_{5c}^0(\Sigma_{5b}^+)$&$-\frac{1}{\sqrt2}$ \\
&&$K^- \Sigma_{5c}^0(\Sigma_{5b}^+)$&$\frac12$&\\
\hline \multicolumn{2}{c}{$\Sigma_{5c}^-(\Sigma_{5b}^0)$} &
\multicolumn{2}{c}{$\Sigma_{5c}^0(\Sigma_{5b}^+)$}&
\multicolumn{2}{c}{$\Theta_{5c}^0(\Theta_{5b}^+)$}
\\ \hline
$\pi^0\Sigma_{5c}^-(\Sigma_{5b}^0)$ & $\frac{1}{2\sqrt2} $ &
$\pi^+\Sigma_{5c}^-(\Sigma_{5b}^0)$ &
$-\frac{1}{2}$ & $K^+\Sigma_{5c}^-(\Sigma_{5b}^0)$ & $\frac{1}{\sqrt2}$  \\
$\pi^-\Sigma_{5c}^0(\Sigma_{5b}^+)$ & $-\frac{1}{2}$ &
$\pi^0\Sigma_{5c}^0(\Sigma_{5b}^+)$ & $-\frac{1}{2\sqrt2}$
& $K^0 \Sigma_{5c}^0(\Sigma_{5b}^+)$ & $-\frac{1}{\sqrt2}$ \\
$\eta_0\Sigma_{5c}^-(\Sigma_{5b}^0)$ & $-\frac{1}{2\sqrt6}$ &
$\eta_0\Sigma_{5c}^0(\Sigma_{5b}^+)$ &
$-\frac{1}{2\sqrt6}$ & $\eta_0 \Theta_{5c}^0(\Theta_{5b}^+)$&$-\frac{2}{\sqrt6}$  \\
$K^+ \Xi_{5c}^{--}(\Xi_{5b}^-)$&$\frac{1}{\sqrt2}$ & $K^+
\Xi_{5c}^-(\Xi_{5b}^0)$ &
$\frac{1}{2}$ && \\
$K^0 \Xi_{5c}^-(\Xi_{5b}^0)$&$-\frac{1}{2}$ &$K^0 \Xi_{5c}^0(\Xi_{5b}^+)$&$-\frac{1}{\sqrt2}$\\
$K^- \Theta_{5c}^0(\Theta_{5b}^+)$&$\frac{1}{\sqrt2}$ &$\bar{K}^0 \Theta_{5c}^0(\Theta_{5b}^+)$&$-\frac{1}{\sqrt2}$\\
\hline
\end{tabular}
\end{center}
\caption{Couplings of the heavy pentaquark anti-sextet $S_{ij}$
with pseudoscalar meson octet $\pi^i_j$. The universal coupling
constant $-\frac{1}{F_\pi}{\cal G}_{S}$ is omitted.}
\label{tab9:SAS}
\end{table}

\begin{table}[h]
\begin{center}
\begin{tabular}{cc|cc|cc}\hline
\multicolumn{2}{c}{$\Sigma_{5c}^{\prime0}(\Sigma_{5b}^{\prime+})$}
&
\multicolumn{2}{c}{$\Sigma_{5c}^{\prime-}(\Sigma_{5b}^{\prime0})$}
& \multicolumn{2}{c}{$\Xi_{5c}^{\prime-}(\Xi_{5b}^{\prime0})$}
\\ \hline
$\pi^+ \Sigma_{5c}^{\prime-}(\Sigma_{5b}^{\prime0})$ & $1$ &
$\pi^0 \Sigma_{5c}^{\prime-}(\Sigma_{5b}^{\prime0})$ &
$-\frac{1}{\sqrt2}$
&$\eta_0 \Xi_{5c}^{\prime-}(\Xi_{5b}^{\prime0})$&$-\frac{2}{\sqrt6}$\\
 $\pi^0 \Sigma_{5c}^{\prime0}(\Sigma_{5b}^{\prime+})$ & $\frac{1}{\sqrt2}$ &
$\eta_0 \Sigma_{5c}^{\prime-}(\Sigma_{5b}^{\prime0})$ &
$\frac{1}{\sqrt{6}}$ &  $\bar{K^0}\Sigma_{5c}^{\prime-}(\Sigma_{5b}^{\prime0})$&$1$\\
$\eta_0 \Sigma_{5c}^{\prime0}(\Sigma_{5b}^{\prime+})$
&$\frac{1}{\sqrt6}$ &
$\pi^-\Sigma_{5c}^{\prime0}(\Sigma_{5b}^{\prime+})$ & $1$
&$K^-\Sigma_{5c}^{\prime0}(\Sigma_{5b}^{\prime+})$&$1$ \\
$K^0 \Xi_{5c}^{\prime-}(\Xi_{5b}^{\prime0})$& $1$ & $K^0
\Xi_{5c}^{\prime-}(\Xi_{5b}^{\prime0})$ & $1$\\
\hline
\end{tabular}
\end{center}
\caption{Couplings of the heavy pentaquark triplet $T^i$ with
pseudoscalar meson octet $\pi^i_j$. The universal coupling
constant ${\cal G}_{TAT}$ is omitted.} \label{tab10:TAT}
\end{table}

\begin{table}
\begin{center}
\begin{tabular}{ccc|ccc|ccc}\hline
\multicolumn{2}{c}{$\Theta^+$} &&
\multicolumn{2}{c}{$\Xi^{--}_{10}$} &&
\multicolumn{2}{c}{$\Theta^0_c(\Theta^+_b)$}& \\
\hline $K^+n$&$-\frac{1}{F_\pi}{\cal C}_{PAB}$&Y& $\pi^-\Xi^-$ &
$-\frac{1}{F_\pi}{\cal C}_{PAB}$ &Y& $D^-(B^0)p$&$-{\cal C}_{SQB}$&Y(Y)\\
$K^0p$&$\frac{1}{F_\pi}{\cal C}_{PAB}$&Y& $K^-\Sigma^-$ &
$\frac{1}{F_\pi}{\cal C}_{PAB}$ &Y& $\bar{D}^0(B^+)n$&${\cal C}_{SQB}$&Y(Y)\\
$K^+N^0_{10}$&$-\frac{1}{\sqrt3}(\frac{1}{F_\pi}{\cal G}_P)$&N&
$\pi^-\Xi^-_{10}$ &$\frac{1}{\sqrt3}(\frac{1}{F_\pi}{\cal G}_P)$&N
 & $D^-(B^0)N^+_{10}$&$-\frac{1}{\sqrt3}{\cal C}_{SQP}$&N(N)\\
 $K^0N^+_{10}$&$\frac{1}{\sqrt3}(\frac{1}{F_\pi}{\cal G}_P)$&N&
$\pi^0\Xi^{--}_{10}$ &$-\frac{1}{\sqrt2}(\frac{1}{F_\pi}{\cal
G}_P)$
 &N& $\bar{D}^0(B^+) N^0_{10}$&$\frac{1}{\sqrt3}{\cal C}_{SQP}$&N(N)\\
$\eta_0 \Theta^+$&$\frac{2}{\sqrt6}(\frac{1}{F_\pi}{\cal G}_P)$&N&
$\eta_0 \Xi^{--}_{10}$ &$-\frac{1}{\sqrt6}(\frac{1}{F_\pi}{\cal
G}_P)$
 &N& $D^-_s(B^0_s) \Theta^+$& ${\cal C}_{SQP}$&N(N)\\
 $K^+n_{8,1}$&$-\frac{1}{F_\pi}{\cal C}_{O_1AP}$&N&
$K^-\Sigma^-_{10}$&$-\frac{1}{\sqrt3}(\frac{1}{F_\pi}{\cal G}_P)$&N&$D^-(B^0)p_{8,1}$&$-{\cal C}_{SQO_1}$&N(N)\\
 $K^0 p_{8,1}$& $\frac{1}{F_\pi}{\cal C}_{O_1AP}$&N&
$\pi^-\Xi^-_{8,1}$ &$-\frac{1}{F_\pi}{\cal C}_{O_1AP}$
 & *& $\bar{D}^0(B^+) n_{8,1}$&${\cal C}_{SQO_1}$&N(N)\\
 $K^+n_{8,2}$&$-\frac{1}{F_\pi}{\cal C}_{O_2AP}$&N&
 $K^-\Sigma^-_{8,1}$ &$\frac{1}{F_\pi}{\cal C}_{O_1AP}$
&N& $D^-(B^0)p_{8,2}$&$-{\cal C}_{SQO_2}$&N(N)\\
 $K^0 p_{8,2}$& $\frac{1}{F_\pi}{\cal C}_{O_2AP}$&N&
$\pi^-\Xi^-_{8,2}$ &$-\frac{1}{F_\pi}{\cal C}_{O_2AP}$&Y& $\bar{D}^0(B^+) n_{8,2}$&${\cal C}_{SQO_2}$&N(N)\\
&&&$K^-\Sigma^-_{8,2}$ &$\frac{1}{F_\pi}{\cal C}_{O_2AP}$
 &*& $K^+\Sigma^-_{5c}$&$-\frac{1}{\sqrt2}(\frac{1}{F_\pi}{\cal G}_S)$&N\\
&&&&&& $K^0\Sigma^0_{5c}$&$\frac{1}{\sqrt2}(\frac{1}{F_\pi}{\cal G}_S)$&N\\
&&&&&& $\eta_0\Theta^0_c$&$\frac{2}{\sqrt6}(\frac{1}{F_\pi}{\cal G}_S)$&N\\
&&&&&&$K^0\Sigma^{\prime0}_{5c}$&$-\frac{1}{F_\pi}{\cal C}_{TAS}$&*\\
&&&&&&$K^+\Sigma^{\prime-}_{5c}$&$\frac{1}{F_\pi}{\cal C}_{TAS}$&*\\
\hline \multicolumn{2}{c}{$\Xi^+_{10}$} &&
\multicolumn{2}{c}{$\Xi^{--}_{5c}(\Xi^{-}_{5b})$} &&
\multicolumn{2}{c}{$\Xi^0_{5c}(\Xi^+_{5b})$}& \\
\hline
$\pi^+\Xi^0$&$-\frac{1}{F_\pi}{\cal C}_{PAB}$&Y&$D^-(B^0)\Xi^-$&${\cal C}_{SQB}$&Y(Y)&$\bar{D}^0(B^+)\Xi^0$&$-{\cal C}_{SQB}$&Y(Y)\\
$\bar{K}^0\Sigma^+$&$\frac{1}{F_\pi}{\cal C}_{PAB}$&Y&$D^-_s(B^0_s)\Sigma^-$&$-{\cal C}_{SQB}$&Y(Y)&$D^-_s(B^0_s)\Sigma^+$&${\cal C}_{SQB}$&Y(Y)\\
$\pi^+\Xi^0_{10}$&$\frac{1}{\sqrt3}(\frac{1}{F_\pi}{\cal G}_P)$&N&$\bar{D}^0(B^+)\Xi^{--}_{10}$&${\cal C}_{SQP}$&N(N)&$\bar{D}^0(B^+)\Xi^0_{10}$&$\frac{1}{\sqrt3}{\cal C}_{SQP}$&N(N)\\
$\pi^0\Xi^+_{10}$&$\frac{1}{\sqrt2}(\frac{1}{F_\pi}{\cal G}_P)$&N&$D^-(B^0)\Xi^-_{10}$&$-\frac{1}{\sqrt3}{\cal C}_{SQP}$&N(N)&$D^-(B^0)\Xi^+_{10}$&$-{\cal C}_{SQP}$&N(N)\\
$\eta_0\Xi^+_{10}$&$-\frac{1}{\sqrt6}(\frac{1}{F_\pi}{\cal G}_P)$&N&$D^-_s(B^0_s)\Sigma^-_{10}$&$\frac{1}{\sqrt3}{\cal C}_{SQP}$&N(N)&$D^-_s(B^0_s)\Sigma^+_{10}$&$\frac{1}{\sqrt3}{\cal C}_{SQP}$&N(N)\\
$\bar{K}^0\Sigma^+_{10}$&$\frac{1}{\sqrt3}(\frac{1}{F_\pi}{\cal G}_P)$&N&$D^-(B^0)\Xi^-_{8,1}$&${\cal C}_{SQO_1}$&N(N)&$\bar{D}^0(B^+)\Xi^0_{8,1}$&$-{\cal C}_{SQO_1}$&N(N)\\
$\pi^+\Xi^0_{8,1}$&$-\frac{1}{F_\pi}{\cal C}_{O_1AP}$&*&$D^-_s(B^0_s)\Sigma^-_{8,1}$&$-{\cal C}_{SQO_1}$&N(N)&$D^-_s(B^0_s)\Sigma^+_{8,1}$&${\cal C}_{SQO_1}$&N(N)\\
$\bar{K}^0\Sigma^+_{8,1}$&$\frac{1}{F_\pi}{\cal C}_{O_1AP}$&N&$D^-(B^0)\Xi^-_{8,2}$&${\cal C}_{SQO_2}$&Y(*)&$\bar{D}^0(B^+)\Xi^0_{8,2}$&$-{\cal C}_{SQO_2}$&Y(*)\\
$\pi^+\Xi^0_{8,2}$&$-\frac{1}{F_\pi}{\cal C}_{O_2AP}$&Y&$D^-_s(B^0_s)\Sigma^-_{8,2}$&$-{\cal C}_{SQO_2}$&Y(Y)&$D^-_s(B^0_s)\Sigma^+_{8,2}$&${\cal C}_{SQO_2}$&Y(Y)\\
$\bar{K}^0\Sigma^+_{8,2}$&$\frac{1}{F_\pi}{\cal C}_{O_2AP}$&*&$\pi^0\Xi^{--}_{5c}(\Xi^{-}_{5b})$&$-\frac{1}{\sqrt2}(\frac{1}{F_\pi}{\cal G}_S)$&N&$\pi^+\Xi^-_{5c}(\Xi^0_{5b})$&$\frac{1}{\sqrt2}(\frac{1}{F_\pi}{\cal G}_s)$&N\\
&&&$\pi^-\Xi^{-}_{5c}(\Xi^0_{5b})$&$\frac{1}{\sqrt2}(\frac{1}{F_\pi}{\cal G}_S)$&N&$\pi^0\Xi^0_{5c}(\Xi^+_{5b})$&$\frac{1}{\sqrt2}(\frac{1}{F_\pi}{\cal G}_s)$&N\\
&&&$\eta_0\Xi^{--}_{5c}(\Xi^{-}_{5b})$&$-\frac{1}{\sqrt6}(\frac{1}{F_\pi}{\cal G}_S)$&N&$\eta_0\Xi^0_{5c}(\Xi^+_{5b})$&$-\frac{1}{\sqrt6}(\frac{1}{F_\pi}{\cal G}_s)$&N\\
&&&$K^-\Sigma^{-}_{5c}(\Sigma^0_{5b})$&$-\frac{1}{\sqrt2}(\frac{1}{F_\pi}{\cal G}_S)$&Y&$\bar{K}^0\Sigma^0_{5c}(\Sigma^+_{5b})$&$\frac{1}{\sqrt2}(\frac{1}{F_\pi}{\cal G}_s)$&Y\\
&&&$K^-\Sigma^{\prime-}_{5c}(\Sigma^{\prime0}_{5b})$&$-\frac{1}{F_\pi}{\cal C}_{TAS}$&Y&$\bar{K}^0\Sigma^{\prime0}_{5c}(\Sigma^{\prime+}_{5b})$&$\frac{1}{F_\pi}{\cal C}_{TAS}$&Y\\
&&&$\pi^-\Xi^{\prime-}_{5c}(\Xi^{\prime0}_{5b})$&$\frac{1}{F_\pi}{\cal C}_{TAS}$&Y&$\pi^+\Xi^{\prime-}_{5c}(\Xi^{\prime0}_{5b})$&$-\frac{1}{F_\pi}{\cal C}_{TAS}$&Y\\
 \hline
\end{tabular}
\end{center}
\caption{Strong decay modes of the three observed pentaquarks and
other exotic pentaquarks with corresponding coupling constants. Y
or N represents the decay mode which is kinematically allowed or
forbidden in JW's model with the masses estimated in Ref.
\cite{jaffe,huang2,wise,zhanga}. Y or N in the parentheses
corresponds to the case of the heavy pseudoscalar meson being
replaced by the heavy vector meson. Whenever the pentaquark lies
very close to the threshold of the final state, we indicate this
case with *.} \label{tab11}
\end{table}

\end{document}